\documentclass[12pt]{article}
\RequirePackage{amsthm,amsmath,amsfonts,amssymb}
\RequirePackage{natbib}

\RequirePackage[colorlinks,citecolor=blue,urlcolor=blue,backref=page,backref=page]{hyperref}

\usepackage{verbatim}
\RequirePackage{mathtools}
\mathtoolsset{showonlyrefs=true}

\usepackage{comment}
\usepackage[inline]{enumitem}

\theoremstyle{plain}
\newtheorem{theorem}{Theorem}[section]
\newtheorem{proposition}[theorem]{Proposition}

\usepackage{graphicx,booktabs,float}
\usepackage{tikz}
\usetikzlibrary{arrows.meta, positioning, calc, fit, shapes.geometric, backgrounds}
\usepackage{algorithm}
\usepackage{algpseudocode}

\usepackage{xcolor}
\newif\ifshowrev
\showrevfalse     % changes shown in red: for coauthor review and resubmission
\newcommand{\rev}[1]{\ifshowrev\textcolor{red}{#1}\else#1\fi}

\begin{document}

\title{Summary Sufficiency in Simulation-Based Inference via Mutual Information}
\author{Renjie Peng, Christian Genest, David A. Stephens\\ \ \\
	Department of Mathematics and Statistics, McGill University, Canada\\
	}
\date{}
\maketitle

\begin{abstract}
Simulation-based inference (SBI) enables posterior estimation when the likelihood is intractable but data can be simulated from the model. For high-dimensional or variable-length data, SBI methods typically rely on low-dimensional summary statistics as conditioning inputs for posterior estimation, making their sufficiency crucial for accurate inference. Mutual information (MI) provides a natural information-theoretic characterization of sufficiency, quantifying how much parameter information a summary retains. However, these MI quantities are generally intractable in likelihood-free settings. \citet{sui2026evaluate} estimate MI from a trained neural posterior using the Barber--Agakov (BA) lower bound, while discussing, but not computing, a Fisher-information upper bound. This makes the estimate ambiguous: a small lower bound may indicate either an insufficient summary or an inaccurate density estimator. We address this limitation with a tractable diagnostic that estimates the BA gap and reveals what the difference between bounds has actually measured. We extend the resulting sufficiency metric to observation-centered restricted data spaces, allowing simulations to focus on a neighborhood of the data while preserving the target posterior and enabling valid sufficiency comparisons between summaries. We illustrate the use of this diagnostic tool on a temporal epidemic-type aftershock sequence model by comparing learned and hand-crafted summary statistics for variable-length earthquake catalogs. 
\end{abstract}


%%%
%%% SECTION 1
%%%

\section{Introduction}

Bayesian inference provides a principled framework for updating a prior distribution \(\pi_0\) for a parameter $\theta \in \Theta$ of interest in light of observed data $x_0 \in \mathcal{X}$ through the posterior distribution, defined for all $\theta \in \Theta$, by
\[
\pi(\theta\mid x_0)\propto p(x_0\mid \theta)\pi_0(\theta),
\]
where $p (x_0 \mid \cdot)$ denotes the likelihood function. In many scientific models, however, this  likelihood is analytically or computationally intractable, making posterior computation difficult. When data can be simulated from the model, simulation-based inference (SBI) addresses this challenge by approximating the posterior using simulated parameter-data pairs \((\theta, X)\) rather than direct likelihood evaluations \citep{cranmer2020frontier,deistler2025simulation}. To emphasize that the subsequent developments hold for any value of $x \in \mathcal{X}$, the capital letter $X$ is used instead, despite the fact that uppercase letters are conventionally reserved for random variables.

In this work, we focus on neural posterior estimation (NPE), a variant of SBI that directly learns a neural posterior estimator $q_{F(X, \phi)}$  approximating $\pi( \cdot \mid X)$ for any $x \in \mathcal{X}$. Here \(F(\cdot,\phi)\) is a neural network with weights \(\phi\), and \(F(X, \phi ) = \psi\) denotes the parameters of a chosen density family \(q_\psi\). Sequential extensions can further improve simulation efficiency for a fixed observation, at the cost of amortization \citep{SNPE-A,SNPE-B,SNPE-C,TSNPE}.

For high-dimensional or variable-length simulator outputs, where the amount of output data generated varies across replications, SBI pipelines commonly compress \(X\) through a summary map \(s\) and condition on the fixed-dimensional summary \(s(X)\) rather than on \(X\) itself \citep{chen2023learning,stockman2024sb}. Ideally, \(s(X)\) is sufficient in the Bayesian sense, that is, conditioning on the summary preserves the posterior based on the full data, i.e., $\pi ( \cdot \mid X) = \pi\{ \cdot \mid s(X)\}$.  However, sufficient summaries are difficult to derive for complex models, making it increasingly important to quantify the parameter information retained by a summary and to compare different summaries on this basis.  Mutual information (MI) offers a natural way to assess summary sufficiency, but computing MI is generally intractable in likelihood-free settings. A recent approach of \citet{sui2026evaluate} addresses this difficulty by estimating MI through the Barber--Agakov (BA) lower bound \citep{BAlowerbound} computed from a neural posterior estimator \(q_{F\{s(X),\phi\}} \); see Section \ref{subsec:ba}. Treating this lower bound as an estimate of MI is reliable only to the extent that the estimator accurately approximates the true summary-conditioned posterior, \(\pi \{ \cdot \mid s(X) \} \). Without checking this condition, a small lower bound for a given summary is ambiguous: it may indicate that the summary loses too much parameter information, or instead that its neural posterior estimator is insufficiently accurate. Although upper bounds could in principle help bracket the true MI, they can be difficult to compute in likelihood-free problems, are common to all summaries, and may be uninformative when loose\rev{; see Section~\ref{subsec:ba}}.

We address this limitation by developing a diagnostic for the gap between the BA lower bound and the corresponding MI. Although this gap involves the unknown summary-conditioned posterior, we show that it can be reformulated as a Kullback--Leibler (KL) divergence between two joint distributions over parameters and summaries that are both amenable to sampling using the simulator and the learned neural posterior estimator. This yields a classifier-based diagnostic for bound tightness and gap differences between summaries. As a result, BA-bound comparisons become interpretable: when the gaps are comparable, differences in lower bounds can be interpreted as differences in MI, and hence in retained parameter information; when they are not, the difference between the bounds cannot in general be read as a difference in MI, and the diagnostic indicates the direction in which the comparison is distorted\rev{; see Section~\ref{subsec:gap-diagnostic}}.

Beyond this diagnostic, we extend MI-based sufficiency comparisons to observation-centered restricted data spaces. Rather than training over the full prior predictive distribution, we restrict simulations to a data-space region \(M\) containing the observation \(x_0\), while preserving the posterior of interest under the restricted joint distribution \rev{(Proposition~\ref{prop:posterior-preservation})}. This restriction keeps MI-based comparisons and BA-gap diagnostics well defined, and improves efficiency for variable-length outputs by concentrating simulation and training on the observed regime. We apply the framework to temporal epidemic-type aftershock sequence (ETAS) earthquake catalogs, i.e., collections of observed earthquake events with their times and magnitudes, comparing learned summaries with hand-crafted summaries inspired by \cite{stockman2024sb}. Separately, we benchmark posterior accuracy against a \rev{Markov chain Monte Carlo} (MCMC) reference targeting the same posterior, a comparison the data-space restriction makes available.

The remainder of the paper is organized as follows. Section~\ref{sec:background} introduces NPE methodology, summary sufficiency, MI, and the BA lower bound. Section~\ref{sec:methodology} develops the BA-gap diagnostic and extends sufficiency comparisons to observation-centered restricted data spaces. Section~\ref{sec:etas} applies the framework to the temporal ETAS model, comparing learned and hand-crafted summaries and benchmarking posterior accuracy against MCMC. Section~\ref{sec:discussion} discusses limitations and concludes.

%%%
%%% SECTION 2
%%%

\section{Background}
\label{sec:background}

%%%
%%% SECTION 2.1
%%%

\subsection{Neural posterior estimation with summaries}
\label{subsec:npe-summary}

We consider a simulator-based model with parameter \(\theta \in \Theta\) and output \(X \in \mathcal{X}\). In the likelihood-free setting, the likelihood \(p(X\mid\theta)\) cannot be evaluated, but a random sample $(\theta_1, X_1), \dots, (\theta_{\rev{N}}, X_{\rev{N}})$ can be generated by drawing successively, for each $i \in \{1, \dots, \rev{N} \}$,
\[
\theta_i \sim \pi_0 \quad \mbox{and} \quad X_i \sim p( \cdot \mid \theta_i).
\]
NPE uses these simulated pairs to train a neural network \(F(\cdot,\phi)\), which defines a conditional density estimator \(q_\psi(\theta) = q_{F(X,\phi)}(\theta)\) for the posterior \(\pi(\theta\mid X)\). 

The population objective underlying NPE \rev{is}
\[
\mathbb{E}_{\pi_0(\theta)p(X\mid\theta)} \left[ \ln \{ q_{F(X,\phi)}(\theta) \} \right]\rev{,}
\]
\rev{to be maximized over \(\phi\). The corresponding estimate \(\hat\phi_N\) minimizes the negative of its sample analog, i.e.,} the empirical loss
%%%
%%% EQUATION (1)
%%%
\begin{equation}
\label{eq:npe-empirical-loss}
\mathcal{L}_N(\phi)
=
-\frac{1}{N}
\sum_{i=1}^N
\ln \{ q_{F(X_i,\phi)}(\theta_i \rev{)} \}.
\end{equation}
When \(F(\cdot,\phi)\) and \(q_\psi\) are
sufficiently expressive, the estimator \(q_{F(X,\hat\phi_N)}\) approximates the posterior \(\pi ( \cdot \mid X)\) for arbitrary $X \in \mathcal{X}$ in the limit as \(N\to\infty\) \citep{SNPE-A}.

For high-dimensional, irregular, or variable-length simulator outputs, it is often impractical to condition the neural posterior estimator directly on \(X\). A common remedy is to introduce hand-crafted summary statistics through a map \(s:\mathcal{X}\to\mathcal{S}\), usually with \(\mathcal{S}\subseteq\mathbb{R}^{d}\) for some integer \(d\) much smaller than the effective dimension of \(X\), and to train the summary-conditioned estimator \(q_{F\{s(X),\phi\}} \) by applying the loss in Eq.~\eqref{eq:npe-empirical-loss} to the summarized data \(s(X_1), \dots, s(X_{\rev{N}}) \). Such summaries are typically designed according to domain expertise.

Rather than hand-crafted summary statistics, one may learn a summary representation from simulations by inserting an embedding network \(s(\cdot,\eta):\mathcal{X}\to\mathcal{S}\) in front of the density-estimator network \(F(\cdot,\phi)\), where \(\eta\) denotes the trainable weights of the embedding network, kept distinct from \(\phi\). This yields the learned-summary estimator \(q_{F\{s(X,\eta),\phi\}} \). The architecture of the embedding network depends on the structure of the simulator output. The two networks are then trained together by minimizing the negative log-density loss in Eq.~\eqref{eq:npe-empirical-loss}, a procedure known as \textit{end-to-end training} \citep{deistler2025simulation}. Further details are given in Section~S1 of the Supplementary Material.

In either case, the estimator targets a posterior conditional on the summary representation, namely \(\pi\{ \cdot \mid s(X)\}\) for hand-crafted summaries or \(\pi\{ \cdot \mid s(X,\eta)\}\) for learned summaries, rather than the full-data posterior \(\pi( \cdot \mid X)\). The two coincide only when the summary is sufficient. How much information a summary retains therefore reflects how far the posterior conditioned on the summary lies from \(\pi( \cdot \mid X)\), and quantifying it provides the basis on which candidate summaries are compared.

%%%
%%% SECTION 2.2
%%%

\subsection{Mutual information and summary sufficiency}
\label{subsec:mi}

MI is a common criterion for measuring the sufficiency of summary
statistics. Recall first that the prior and the likelihood induce the joint density on \(\Theta\times\mathcal{X}\), which admits the following two factorizations for all $\theta \in \Theta$ and $x \in \mathcal{X}$:
%%%
%%% EQUATION (2)
%%%
\begin{equation}
\label{eq:joint}
\pi_0(\theta)\,p(x\mid\theta)=\pi(\theta\mid x)\,m(x),
\quad
m(x)=\int_\Theta p(x\mid\theta)\,\pi_0(\theta)\,\mathrm{d}\theta ,
\end{equation}
where \(m\) is the prior predictive density. The MI between
\(\theta\) and \(X\) is defined as the KL divergence between this joint density and the product of its marginals, namely
%%%
%%% EQUATION (3)
%%%
\begin{equation}
\label{eq:mi-def}
\begin{split}
I(\pi,\pi_0)
&=
D_{\mathrm{KL}}\bigl\{\pi_0(\theta)\,p(x\mid\theta)\,\big\|\,\pi_0(\theta)\,m(x)\bigr\}\\[2pt]
&=
\int_{\mathcal{X}}\!\int_{\Theta}
\ln\!\left\{\frac{\pi_0(\theta)\,p(x\mid\theta)}{\pi_0(\theta)\,m(x)}\right\}
\pi_0(\theta)\,p(x\mid\theta)\,\mathrm{d}\theta\,\mathrm{d}x\\[2pt]
&=
\int_{\mathcal{X}}\!\int_{\Theta}
\ln\!\left\{\frac{\pi(\theta\mid x)}{\pi_0(\theta)}\right\}
\pi_0(\theta)\,p(x\mid\theta)\,\mathrm{d}\theta\,\mathrm{d}x\\[2pt]
&=
\int_{\mathcal{X}} D_{\mathrm{KL}}\bigl\{\pi(\theta\mid x)\,\big\|\,\pi_0(\theta)\bigr\}\,
m(x)\,\mathrm{d}x ,
\end{split}
\end{equation}
where the third equality follows by substituting Eq.~\eqref{eq:joint} into the integrand. The last line shows that \(I(\pi,\pi_0)\) is the expected prior-to-posterior information gain: it is large when observing \(X\) typically moves the posterior far from the prior, and vanishes when \(\theta\) and \(X\) are independent. The last line in Eq.~\eqref{eq:mi-def} also shows that the mutual information is a function of the posterior and the prior, rather than of \(\theta\) and \(X\) themselves. We therefore write it as \(I(\pi,\pi_0)\) instead of the more common \(I(\theta;X)\). This notation will prove to be convenient later, when the prior is held fixed and each summary map induces its own posterior.

The same construction applies to a summary map \(s\). Drawing \(\theta\sim\pi_0\), simulating \(X\sim p(\cdot\mid\theta)\) and retaining only \(s(X)\) defines a Bayesian model on \(\Theta\times\mathcal{S}\) with the same structure as Eq.~\eqref{eq:joint}. This model is defined, for all $\theta \in \Theta$ and $s \in \mathcal{S}$, by
\begin{equation}
\label{eq:joint-summary}
\pi_0(\theta)\,p_s(s\mid\theta)=\pi(\theta\mid s)\,m_s(s),
\qquad
m_s(s) = \int_\Theta p_s(s\mid\theta)\,\pi_0(\theta)\,\mathrm{d}\theta ,
\end{equation}
where \(p_s(\cdot\mid\theta)\) is the density of \(s(X)\) when \(X\sim p(\cdot\mid\theta)\) and \(\pi( \cdot \mid s)\) is the posterior obtained from observing the summary alone. Pulling this posterior back to the data space, we write
%%%
%%% EQUATION (4)
%%%
\begin{equation}
\label{eq:summary-posterior}
\pi_s (\theta \mid X) = \pi\{\theta\mid s(X) \},
\end{equation}
so that \(\pi_s(\cdot\mid X)\) is the posterior an analyst would report after seeing
\(s(X)\) rather than \(X\) itself. Replacing \(X\) by \(s(X)\) in
Eq.~\eqref{eq:mi-def} then gives
\begin{equation}
\label{eq:mi-summary}
\begin{split}
I(\pi_s,\pi_0)
&=
D_{\mathrm{KL}}\bigl\{\pi_0(\theta)\,p_s(s\mid\theta)\,\big\|\,\pi_0(\theta)\,m_s(s)\bigr\}\\[2pt]
&=
\int_{\mathcal{S}}\!\int_{\Theta}
\ln\!\left\{\frac{\pi(\theta\mid s)}{\pi_0(\theta)}\right\}
\pi_0(\theta)\,p_s(s\mid\theta)\,\mathrm{d}\theta\,\mathrm{d}s\\[2pt]
&=
\int_{\mathcal{X}}\!\int_{\Theta}
\ln\!\left\{\frac{\pi_s(\theta\mid x)}{\pi_0(\theta)}\right\}
\pi_0(\theta)\,p(x\mid\theta)\,\mathrm{d}\theta\,\mathrm{d}x ,
\end{split}
\end{equation}
where the last equality holds because the integrand depends on \(x\) only through \(s(x)\). This final form is the one used in practice: it is an expectation with respect to the simulated pairs \((\theta_1, X_1), \dots, (\theta_{\rev{N}}, X_{\rev{N}}) \), and therefore requires no knowledge of the summary densities \(p_s\) and \(m_s\). Learned summaries are covered by the same expressions upon setting \(s=s(\cdot,\eta)\), in which case we use the abbreviations \(\pi_\eta\) and \(I(\pi_\eta,\pi_0)\) for the corresponding posterior and MI value, respectively.

MI characterizes Bayesian sufficiency exactly. Given that the sequence \(\theta\to X\to s(X)\) forms a Markov chain, the data-processing inequality
\citep[Sec.~2.8]{cover2006elements} yields
%%%
%%% EQUATION (5)
%%%
\begin{equation}
\label{eq:dpi}
I(\pi_s,\pi_0)\le I(\pi,\pi_0),
\end{equation}
with equality if and only if \(\pi_s( \cdot \mid X) = \pi( \cdot \mid X)\) for \(m\)-almost every \(X\), i.e., if and only if \(s\) is Bayes sufficient \citep[Sec.~2.10]{cover2006elements}.  Comparisons between two summaries \(s_1\) and \(s_2\) can therefore be phrased directly in terms of mutual information: \(s_2\) retains more parameter information than \(s_1\) precisely when
%%%
%%% EQUATION (6)
%%%
\begin{equation}
\label{eq:mi-comparison}
I(\pi_{s_1},\pi_0)<I(\pi_{s_2},\pi_0).
\end{equation}

In the likelihood-free setting, however, \(I(\pi_s,\pi_0)\) cannot be evaluated, in view of the fact that the summary-conditioned posterior \(\pi_s\) appearing in Eq.~\eqref{eq:mi-summary} is itself unavailable. To address this, \citet{sui2026evaluate} replace \(I(\pi_s,\pi_0)\) by a computable lower bound obtained from a trained neural posterior estimator. Section~\ref{subsec:ba} describes this bound and identifies the price paid for the substitution.

%%%
%%% SECTION 2.3
%%%

\subsection{The Barber--Agakov lower bound and its gap}
\label{subsec:ba}

A trained neural posterior estimator provides a computable surrogate for the unknown \(\pi_s\). Writing
\begin{equation}
\label{eq:q-shorthand}
q_{s,\phi}(\theta \mid X) = q_{F\{s(X),\phi\}}(\theta),
\end{equation}
which, like \(\pi_s\), depends on \(X\) only through \(s(X)\), and splitting the logarithm in the last line of Eq.~\eqref{eq:mi-summary} as
\[
\ln\!\left\{\frac{\pi_s(\theta\mid x)}{\pi_0(\theta)}\right\}
=
\ln\!\left\{\frac{\pi_s(\theta\mid x)}{q_{s,\phi}(\theta\mid x)}\right\}
+
\ln\!\left\{\frac{q_{s,\phi}(\theta\mid x)}{\pi_0(\theta)}\right\}
\]
yields the decomposition
%%%
%%% EQUATION (7)
%%5
\begin{equation}
\label{eq:ba-decomposition}
I(\pi_s,\pi_0)=\hat I(q_{s,\phi},\pi_0)+\Delta(s,\phi),
\end{equation}
where
%%%
%%% EQUATION (8)
%%5
\begin{equation}
\label{eq:ba-bound}
\hat I(q_{s,\phi},\pi_0)
=
\int_{\mathcal{X}}\!\int_{\Theta}
\ln q_{s,\phi}(\theta\mid x)\,
\pi_0(\theta)\,p(x\mid\theta)\,\mathrm{d}\theta\,\mathrm{d}x
+h(\pi_0),
\end{equation}
is the Barber--Agakov (BA) lower bound, with \(h(\pi_0)=-\int_\Theta\ln\{\pi_0(\theta)\}\pi_0(\theta)\,\mathrm{d}\theta\) the differential entropy of the prior, and
%%%
%%% EQUATION (9)
%%5
\begin{equation}
\label{eq:ba-gap}
\begin{split}
\Delta(s,\phi)
&=
\int_{\mathcal{X}}\!\int_{\Theta}
\ln\!\left\{\frac{\pi_s(\theta\mid x)}{q_{s,\phi}(\theta\mid x)}\right\}
\pi_0(\theta)\,p(x\mid\theta)\,\mathrm{d}\theta\,\mathrm{d}x\\[2pt]
&=
\int_{\mathcal{S}}
D_{\mathrm{KL}}\bigl\{\pi(\theta\mid s)\,\big\|\,q_{F(s,\phi)}(\theta)\bigr\}\,
m_s(s)\,\mathrm{d}s
\end{split}
\end{equation}
is the expected divergence between the true and the estimated summary-conditioned posteriors\rev{, referred to hereafter as the BA gap, or simply the gap}. Since \(\Delta(s,\phi)\ge 0\), Eq.~\eqref{eq:ba-decomposition} gives
\begin{equation}
\label{eq:ba-inequality}
\hat I(q_{s,\phi},\pi_0)\ \le\ I(\pi_s,\pi_0),
\end{equation}
with equality if and only if \(q_{F(s,\phi)} (\cdot ) = \pi( \cdot \mid s)\) for \(m_s\)-almost every \(s\).

Unlike \(I(\pi_s,\pi_0)\), both terms in Eq.~\eqref{eq:ba-bound} can be estimated by Monte Carlo, the first from the simulated pairs and the second from prior draws alone. In practice, the entropy need not be evaluated at all, since it depends only on the prior and cancels when two summaries are compared under a common prior. Comparing Eq.~\eqref{eq:ba-bound} with Eq.~\eqref{eq:npe-empirical-loss} further shows that this estimate is, up to that constant, the negative of the NPE training loss, so no procedure beyond fitting the estimator is required.  For a learned summary, \(s(\cdot,\eta)\), end-to-end training maximizes the BA lower bound over both the embedding and estimator weights \((\eta,\phi)\), rather than the MI itself. The two differ by the corresponding gap; the decomposition is given in Section~S1 of the Supplementary Material.

The bound, however, records the MI only up to \(\Delta(s,\phi)\), and the two cannot be separated from the reported value alone. Subtracting two instances of Eq.~\eqref{eq:ba-decomposition} gives
%%%
%%% EQUATION (10)
%%5
\begin{equation}
\label{eq:gap-difference}
I(\pi_{s_2},\pi_0)-I(\pi_{s_1},\pi_0)
=
\bigl\{\hat I(q_{s_2,\phi_2},\pi_0)-\hat I(q_{s_1,\phi_1},\pi_0)\bigr\}
+\bigl\{\Delta(s_2,\phi_2)-\Delta(s_1,\phi_1)\bigr\},
\end{equation}
so an observed difference between the bounds reproduces the comparison \eqref{eq:mi-comparison} only up to the difference between the gaps. A smaller bound may therefore reflect an insufficient summary or merely an inaccurate density estimator.

\citet{sui2026evaluate} proceed under the assumption that the trained estimator is accurate enough for \(\Delta(s,\phi)\) to be negligible, and report \(\hat I(q_{s,\phi},\pi_0)\) as an estimate of \(I(\pi_s,\pi_0)\). They identify the gap as the bias of this estimate, and note that it is generally infeasible to evaluate because \(\pi(\cdot \mid s)\) is unknown. As an alternative, they discuss an Efromovich-type upper bound involving the prior information and the prior-averaged Fisher information, which is costly to evaluate and which they leave to future work. This is an upper bound on the full-data mutual information \(I(\pi, \pi_0)\), and hence, by Eq.~\eqref{eq:dpi}, also an upper bound on \(I(\pi_s,\pi_0)\) for any summary \(s\). Because it depends on the model and the prior rather than on the summary, however, it takes the same value for every summary, and whenever it is loose, it tells little about how close a given summary is to the full-data mutual information. Therefore, it does not resolve the summary-specific ambiguity caused by the BA gap.

We instead estimate \(\Delta(s,\phi)\) itself. Although Eq.~\eqref{eq:ba-gap} involves the unknown posterior distribution \(\pi(\cdot \mid s)\), Section~\ref{sec:methodology} rewrites it as a divergence between two joint distributions over the pair \((\theta, s(X))\) from which samples are available.

%%%
%%% SECTION 3
%%%

\section{Methodology}
\label{sec:methodology}

%%%
%%% SECTION 3.1
%%%

\subsection{A classifier-based diagnostic for the Barber--Agakov gap}
\label{subsec:gap-diagnostic}

The obstacle to evaluating \(\Delta(s,\phi)\) in Eq.~\eqref{eq:ba-gap} is the unknown posterior \(\pi( \cdot \mid s )\). It disappears once the gap is read as a divergence between two joint distributions rather than as an expected conditional divergence. For arbitrary $\theta \in \Theta$ and $s \in \mathcal{S}$, let
\begin{equation}
\label{eq:two-joints}
P_1(\theta,s) = \pi(\theta\mid s)\,m_s(s),
\qquad
P_2(\theta,s)=q_{F(s,\phi)}(\theta)\,m_s(s)
\end{equation}
be the two densities on \(\Theta\times\mathcal{S}\) obtained by pairing the summary marginal with the true and the estimated conditional, respectively. These distributions share the marginal \(m_s\) and differ only in their conditionals, so the second line of Eq.~\eqref{eq:ba-gap} states precisely that
\begin{equation}
\label{eq:gap-as-joint-kl}
\Delta (s, \phi) = D_{\mathrm{KL}}(P_1\,\|\,P_2).
\end{equation}
Neither density can be evaluated, but both can be sampled, and from the same simulator, as follows. Draw \(n\) pairs \( (\theta_1, X_1), \dots, (\theta_n, X_n) \) from each of $P_1$ and $P_2$ as in Section~\ref{subsec:npe-summary}, and set \(s_1 = s(X_1), \dots, s_n = s(X_n) \). The first \(n\) pairs are a random sample from \(P_1\). For the remaining \(n\) pairs, the parameters $\theta_1, \dots, \theta_n$ are discarded and replaced by draws \(\theta_1'\sim q_{F(s_1, \phi)}, \dots, \theta_n'\sim q_{F(s_n, \phi)}\) from the trained estimator. 

Given that for each integer $i \in \{1 , \dots, n \}$, one has \(s_i \sim m_s\) and \(\theta_i'\) follows the estimated conditional at \rev{\(s_i\)}, the pairs \( (\theta_1', s_1), \dots, (\theta_n', s_n) \) form a random sample from \(P_2\). We then train a binary classifier \(d:\Theta\times\mathcal{S}\to(0,1)\) to distinguish a balanced sample from \(P_1\) and \(P_2\) by minimizing the binary cross-entropy. Its optimum supports two distinct interpretations.

%%%
%%% PROPOSITION 3.1
%%%

\begin{proposition}
\label{prop:two-readouts}
Suppose that the distribution \(P_1\) is absolutely continuous with respect to the distribution \(P_2\). Let \(Y\) be equally likely to be \(0\) or \(1\), and let the pair \((\theta,s)\) be drawn from \(P_1\) when \(Y=1\) and from \(P_2\) when \(Y=0\). Among all measurable maps \(d:\Theta\times\mathcal{S}\to(0,1)\), the binary cross-entropy for predicting \(Y\) from \((\theta,s)\) is minimized at \(d^\star = P_1/(P_1+P_2)\). For this specific value of  \(d^\star\), the following statements hold true.
\begin{itemize}
\item [(a)]
First, the rule \(\hat Y=\mathbf{1}\{d^\star(\theta,s)>1/2\}\) satisfies
%%%
%%% EQUATION (11)
%%%
\begin{equation}
\label{eq:tv-identity}
\Pr(\hat Y=Y)=\tfrac12+\tfrac12\,\mathrm{TV}(P_1,P_2),
\end{equation}
where \(\mathrm{TV}(P_1,P_2) = \tfrac12\int_{\Theta\times\mathcal{S}} |P_1 - P_2|\,\mathrm{d}\theta\,\mathrm{d}s\) is the total variation distance.
\item [(b)]
%%%
%%% EQUATION (12)
%%%
\begin{equation}
\label{eq:kl-identity}
\Delta(s,\phi)
=
\int_{\Theta\times\mathcal{S}}
\ln\!\left\{\frac{d^\star(\theta,s)}{1-d^\star(\theta,s)}\right\}
P_1(\theta,s)\,\mathrm{d}\theta\,\mathrm{d}s .
\end{equation}
\end{itemize}
\end{proposition}

Both parts of Proposition~\ref{prop:two-readouts} are classical results on classifier-based estimation of distances between distributions \citep{lecam1986asymptotic,menon2016linking,nguyen2010estimating}; a proof is given in Section~S2 of the Supplementary Material.  Only Eq.~\eqref{eq:kl-identity} is useful for our purpose. Classification accuracy estimates the total variation distance, not the divergence \(\Delta(s,\phi)\) appearing in Eq.~\eqref{eq:ba-decomposition}; the two vanish together, so an accuracy near \(1/2\) does certify a tight bound, but away from zero they are not in monotone correspondence and accuracy cannot supply the correction in Eq.~\eqref{eq:gap-difference}. Replacing \(d^\star\) by a fitted classifier \(\hat d\) and the integral by an average over held-out draws from \(P_1\) gives
\begin{equation}
\label{eq:gap-estimator}
{\hat \Delta} =
\frac{1}{n}\sum_{i=1}^{n}
\ln\!\left\{\frac{\hat d(\theta_i,s_i)}{1-\hat d(\theta_i,s_i)}\right\},
\end{equation}
where $(\theta_1, s_1), \dots, (\theta_n, s_n)$ are mutually independent draws from $P_1$. Furthermore, ${\hat \Delta}$ is a consistent estimate of $\Delta (s, \phi)$ provided \(\hat d\to d^\star\).

Comparing the sufficiency of two summaries then proceeds in two steps. 
\begin{enumerate}
\item
The first step is the one taken by
\citet{sui2026evaluate}: compute the difference between the two BA lower bounds,
which is available directly from the trained estimators. 
\item 
The second step is the tightness test proposed here. Fitting a classifier for each summary gives \(\hat\Delta_1\) and
\(\hat\Delta_2\), and Eq.~\eqref{eq:gap-difference} then determines what the first step has actually measured. 
\end{enumerate}

When the two gaps are comparable, the difference between the bounds is an estimate of the difference in mutual information; when they are not, the sign of \(\hat\Delta_2-\hat\Delta_1\) says whether that difference understates or overstates it, and its magnitude informs as to whether the ordering survives. Since \(\hat\Delta\) is possibly biased for finite sample size \(n\), we apply the same procedure to two disjoint halves of a single sample from \(P_1\), for which the divergence is zero by construction. The resulting values serve as a reference: they reveal any systematic offset introduced by the classifier, and their spread fixes the scale below which an estimated gap cannot be told from zero. We recommend reporting this reference alongside any application of the diagnostic; it costs no additional
simulation.

%%%
%%% SECTION 3.2
%%%

\subsection{Observation-centered restricted data spaces}
\label{subsec:restricted}

Training the neural posterior over the full prior predictive may be wasteful when the observation \(x_0 \) lies in a comparatively small region of \(\mathcal{X}\). Restricting the simulations to a region containing \(x_0 \) is possible because, as Proposition~\ref{prop:posterior-preservation} below shows, it leaves the target \(\pi( \cdot \mid x_0 )\) unchanged. It is also worthwhile for the following two reasons.

\begin{itemize}
\item [(a)]
\textbf{Cost of simulation:} For models with a branching structure, such as the temporal ETAS model considered in Section~\ref{sec:etas}, the prior may place mass on parameter values for which the process is supercritical and the simulated output can contain arbitrarily many events. A data-space restriction based on catalog size can control this cost by imposing an upper bound on the number of events and terminating the simulation once the bound is exceeded.
\item [(b)]
\textbf{Relevance:} excluding data unlike \(x_0\) does not deprive the sufficiency diagnostic of its meaning. The MI under the restriction still measures how much parameter information a summary retains, but now within the regime to which the observation belongs, which is more relevant to the target inference.
\end{itemize}

Concerning simulation cost, one possible remedy is to confine the prior to a relevant portion of the parameter space. Doing so, however, may introduce a joint constraint on the parameter space, which creates an additional difficulty for neural posterior estimation. When the prior support is bounded, a neural density estimator with unbounded support may assign non-negligible probability mass outside the prior support, a phenomenon commonly referred to as mass leakage. For supports of product form, such as boxes, this issue can often be avoided by mapping each coordinate to the whole real line through a fixed invertible transformation, training in the unconstrained space, and mapping posterior draws back afterwards. 
By contrast, when the support is defined by a joint nonlinear constraint, no analogous componentwise transformation is generally available. Rejection sampling is an alternative solution, but the acceptance rate depends on how much of the posterior mass falls outside the constraint, so drawing a fixed number of posterior samples can be notoriously wasteful. The data-space restriction considered here avoids imposing such a constraint on the prior.

We therefore fix a measurable set \(M\subseteq\mathcal{X}\) with \(x_0 \in M\) and \(\mathbb{P}(M) = \int_M m(x)\,\mathrm{d}x > 0\), and retain only simulated pairs whose data component falls in \(M\). The set is defined through the drawn value of \(X\) alone: it does not depend on \(\theta\) or on the summary map.
Conditioning the joint distribution of the pair $(\theta, X)$ on the event \(\{X\in M\}\) leads to the restricted joint density defined, for all $\theta \in \Theta$ and $x \in \mathcal{X}$, by
%%%
%%% EQUATION (14)
%%%
\begin{equation}
\label{eq:restricted-joint}
p^M(\theta,x) = \mathbf{1}(x\in M)\,\frac{\pi_0(\theta)\,p(x\mid\theta)}{\mathbb{P}(M)} ,
\end{equation}
where \rev{\(\mathbf{1}(\cdot)\) equals \(1\) when the condition in its argument holds and \(0\) otherwise}.

Both marginals differ from their unrestricted counterparts. Integrating Eq.~\eqref{eq:restricted-joint} over \(\theta\), and using the fact that the indicator and \(\mathbb{P}(M)\) are constant, one finds that the prior predictive truncated to \(M\) is given, for all $x \in \mathcal{X}$, by
%%%
%%% EQUATION (15)
%%%
\begin{equation}
\label{eq:restricted-x-marginal}
m^M(x) = \mathbf{1} (x\in M) \,\frac{m(x)}{\mathbb{P}(M)} ,
\end{equation}
while integrating over \(x\) yields the reweighted prior given, for all $\theta \in \Theta$, by
\begin{equation}
\label{eq:restricted-prior}
\pi_0^M(\theta)
=
\pi_0(\theta)\,\frac{\mathbb{P}(M\mid\theta)}{\mathbb{P}(M)},
\qquad
\mathbb{P}(M\mid\theta) = \int_M p(x\mid\theta)\,\mathrm{d}x .
\end{equation}
Here \(\mathbb{P}(M\mid\theta)\) is the probability that \(\theta\) produces data in \(M\), and \(\mathbb{P}(M)\) is its prior average, so parameter values more likely than average to produce data in \(M\) are up-weighted while the others are down-weighted; and \(\pi_0^M\) remains proper, because \(\int_\Theta\pi_0(\theta)\,\mathbb{P}(M\mid\theta)\,\mathrm{d}\theta = \mathbb{P}(M)\).

Equation~\eqref{eq:restricted-joint} defines a Bayesian model on \(\Theta\times\mathcal{X}\) of exactly the form treated in Section~\ref{subsec:mi}, with prior \(\pi_0^M\) and prior predictive \(m^M\). Every quantity of Sections~\ref{subsec:mi}--\ref{subsec:gap-diagnostic} therefore carries over. We write \(\pi^M( \cdot \mid x)\) for the posterior, \(\pi_s^M\) for the summary-conditioned posterior and \(q^M_{s,\phi}\) for the estimator, all under the restricted joint, so that the restricted MI and BA lower bounds are
\[
I(\pi_s^M,\pi_0^M) \quad \mbox{and} \quad \hat I(q^M_{s,\phi},\pi_0^M), 
\]
respectively. The corresponding gap, whose arguments are the summary map and the estimator weights rather than distributions, is written \(\Delta^M(s,\phi)\). Rejection produces exact draws from Eq.~\eqref{eq:restricted-joint} without ever evaluating \( \mathbb{P}(M\mid\cdot )\), so the diagnostic of Section~\ref{subsec:gap-diagnostic} applies \textit{verbatim} with all samples drawn under the restricted joint.

%%%
%%% PROPOSITION 3.2
%%%

\begin{proposition}
\label{prop:posterior-preservation}
For every \(x\in M\) \rev{with \(m(x)>0\)}, \(\pi^M (\cdot \mid x) = \pi (\cdot \mid x)\). In particular \(\pi^M ( \cdot \mid x_0) = \pi ( \cdot \mid x_0)\), so the inference target is unchanged.
\end{proposition}

\begin{proof}
Fix \(x\in M\) \rev{with \(m(x)>0\)}. By definition, \rev{
\(\pi^M (\cdot \mid x) = p^M(\cdot,x)/m^M(x) \)}. Dividing
Eq.~\eqref{eq:restricted-joint} by Eq.~\eqref{eq:restricted-x-marginal}, the
indicator and the factor \(\mathbb{P}(M)\) cancel, so that, by Eq.~\eqref{eq:joint}, one has
\[
\pi^M(\cdot \mid x) = \frac{\pi_0 (\cdot)\, p(x \mid \cdot )}{m(x)} = \pi(\cdot \mid x).
\]
This completes the argument.
\end{proof}

Training on restricted simulations minimizes
\[
\mathbb{E}_{m^M(X)}[D_{\mathrm{KL}}\{\pi^M(\theta\mid X)\,\|\, q^M_{s,\phi}(\theta\mid X)\}], 
\]
which by Proposition~\ref{prop:posterior-preservation} equals \(\mathbb{E}_{m^M(X)}[D_{\mathrm{KL}}\{\pi(\theta \mid X)\,\|\, q^M_{s,\phi}(\theta\mid X)\}]\). The restriction therefore enters only through the distribution over \(X\) on which the loss is averaged. The estimator should still match \(\pi( \cdot \mid X)\) at every \(X\in M\), but the training data now concentrate on data resembling \(x_0\), so the fit is sharpened there. 

Because \(M\) is defined through \(X\) alone, \(\pi_0^M\) is the same for every summary map. The restricted analog of Eq.~\eqref{eq:gap-difference} is therefore
%%%
%%% EQUATION (16)
%%%
\begin{multline}
\label{eq:restricted-gap-difference}
I(\pi_{s_2}^M,\pi_0^M)-I(\pi_{s_1}^M,\pi_0^M)
\\ =
\bigl\{\hat I(q^M_{s_2,\phi_2},\pi_0^M)-\hat I(q^M_{s_1,\phi_1},\pi_0^M)\bigr\}
+\bigl\{\Delta^M(s_2,\phi_2)-\Delta^M(s_1,\phi_1)\bigr\},
\end{multline} 
in which \(h(\pi_0^M)\) has canceled between the two bounds. This matters more than in the unrestricted case: unlike \(h(\pi_0)\), the restricted entropy admits no closed form and cannot be evaluated by sampling, since \(\pi_0^M\) involves the intractable factor \(\mathbb{P}(M\mid\theta)\). Working with differences is therefore not merely convenient but necessary.

The restricted MI nonetheless differs in value from its unrestricted counterpart, measuring what a summary retains within the narrower regime containing the observation. In this respect, the construction is a regional analog of the L-C2ST diagnostic \citep{linhart2023lc2st}, which localizes posterior-approximation diagnostics at a single \(x_0\); the choice of \(M\) is discussed further in Section~\ref{sec:discussion}.

%%%
%%% SECTION 4
%%%

\section{Application to the Temporal ETAS Model}
\label{sec:etas}

%%%
%%% SECTION 4.1
%%%

\subsection{The temporal ETAS model and MCMC reference}
\label{subsec:etas-setup}

The temporal ETAS model \citep{ogata1988} is a marked Hawkes process for catalogs \(X = \{(t_1, m_1), \dots, (t_{n(X)}, m_{n(X)} ) \} \subset[0,T] \times [M_0,\infty)\), recording the random times of earthquakes together with their magnitudes. It is specified by its conditional intensity function (CIF), \(\lambda(t,m\mid\mathcal{H}_t)\), which represents the expected rate of earthquakes at time \(t\) with magnitude \(m\), given the observation history \(\mathcal{H}_t\) up to time $t$. In the standard formulation, the CIF is given, for any $t \in [0, T]$ and $m \in [M_0, \infty)$, by
\begin{equation}
\label{eq:etas-intensity}
\lambda(t,m\mid\mathcal{H}_t)
=
\Bigl\{\mu+\sum_{j:t_j<t} g(t-t_j,m_j)\Bigr\}\,f_{\mathrm{GR}}(m),
\end{equation}
where \(\mu\) is a constant background rate of events, \(g\) is a non-negative excitation kernel describing how past events contribute to the rate of future ones, and \(f_{\mathrm{GR}}\) is the density of the magnitude. Here, \(g\) factorizes into a magnitude term and an elapsed-time term. That is, for every pair $(t, m) \in [0, T] \times [M_0, \infty)$, one has 
\begin{equation}
\label{eq:etas-kernel}
g(t,m) = k(m;K,\alpha)\,h(t;c,p).
\end{equation}
The magnitude contribution is governed by the Utsu law of productivity, which specifies that, for every $m \in [M_0, \infty)$, 
%%%
%%% EQUATION (17)
%%%
\begin{equation}
\label{eq:etas-utsu}
k(m;K,\alpha) = K\,\mathrm{e}^{\alpha(m-M_0)}.
\end{equation}
The temporal decay follows the Omori--Utsu decay law, which states that for each time $t \in [0, \infty)$,
\begin{equation}
\label{eq:etas-omori}
h(t;c,p) = c^{\,p-1}(p-1)(t+c)^{-p}.
\end{equation}
Magnitudes are assumed to be unpredictable, in the sense that they do not depend on previous events; they are drawn independently from the Gutenberg--Richter law defined, for all $m \in [M_0, \infty)$, by
\begin{equation}
\label{eq:etas-gr}
f_{\mathrm{GR}}(m)=\beta_m\,\mathrm{e}^{-\beta_m(m-M_0)}.
\end{equation}

In the following experiment, we fix \(T=1000\), \(M_0=3\) and \(\beta_m = \ln (10) \approx 2.3 \), corresponding to a \(b\)-value of \(1\) in the Gutenberg--Richter relation. The model then comprises five parameters of interest which can be combined in the vector \(\theta = (\mu, K, \alpha, c, p) \in \Theta = (0, \infty)^4 \times (1, \infty) \).

Given a catalog \(x_0\) of \(n = n(x_0)\) events observed on \([0,T]\), the log-likelihood is given, for any $\theta \in \Theta$, by
\begin{equation}
\label{eq:etas-loglik}
\begin{split}
\ln \{ p(x_0\mid\theta) \}
& =
\sum_{i=1}^{n}
\ln\Bigl\{\mu+\sum_{j<i} k(m_j;K,\alpha)\,h(t_i-t_j;c,p)\Bigr\}
-\mu T\\[2pt]
&\quad
-\sum_{i=1}^{n} k(m_i;K,\alpha)\,H(T-t_i;c,p)
+\sum_{i=1}^{n}\ln f_{\mathrm{GR}}(m_i),
\end{split}
\end{equation}
where
\[
H(\rev{\tau};c,p) = \int_0^{\rev{\tau}} h(u;c,p)\,\mathrm{d}u = 1-\{c/(\rev{\tau}+c)\}^{\, p-1}
\]
is the integrated Omori--Utsu kernel. Since \(\beta_m\) is fixed, the last term does not involve \(\theta\) and plays no part in inference.

The posterior \(\pi( \cdot \mid x_0) \propto p(x_0 \mid \cdot )\pi_0( \cdot )\) has no closed form, and we sample from it with the MCMC scheme of \citet{ross2021}, which introduces additional latent variables to break the dependence between the parameters in the likelihood, as implemented in the \texttt{bayesianETAS} package of \citet{ross2021}. Evaluating the likelihood requires a double summation over event pairs, so its cost grows quadratically in the number of events, which limits such samplers to moderate catalogs \citep{stockman2024sb}. Simulation is far cheaper: generating a catalog from the branching representation \citep{zhuang2004} requires a single pass over events followed by a chronological sort, which motivates the application of SBI. Here, however, the observation window \(T = 1000\) keeps catalogs small enough for MCMC to remain practical. This will allow us, in Section~\ref{subsec:etas-c2st}, to check whether the posterior based
on the summary favored by our diagnostic is closer to the MCMC reference than the posterior based on the alternative summary.

We assign a Gamma prior to the background rate, concentrating its mass on small positive values, and independent uniform priors to the remaining parameters over ranges chosen to cover their plausible physical values, based on earlier work \citep{Dutilleul/Genest/Peng:2024, Peng/Dutilleul/Genest:2025}:
%%%
%%% EQUATION (18)
%%%
\begin{equation}
\label{eq:etas-prior}
\begin{gathered}
\mu\sim \mathcal{G} (\text{shape}=1,\ \text{rate}=5),
\qquad
K\sim \mathcal{U} (0.01,0.50),
\qquad
\alpha\sim \mathcal{U} (0.5,1.5),\\[2pt]
c\sim \mathcal{U} (0.005,0.05),
\qquad
p\sim \mathcal{U} (1.05,2.0).
\end{gathered}
\end{equation}
The Gamma prior on \(\mu\) is the exponential distribution with mean \(0.2\). This prior is not confined to the `subcritical' region \citep{stockman2024sb}: the branching ratio \(K\beta_m/(\beta_m-\alpha)\) exceeds one on part of its support, so a fraction of draws generate explosive catalogs. These are handled using the data-space restriction of Section~\ref{subsec:restricted}, which we define here through catalog size. Writing \(n(x)\) for the number of events in the catalog corresponding to some $x \in \mathcal{X}$, we take
\begin{equation}
\label{eq:etas-restriction}
M=
\bigl\{
x\in\mathcal{X}:
\lfloor n(x_0)/4\rfloor
\le n(x)
\le 4\,n(x_0)
\bigr\},
\end{equation}
whose upper bound caps the simulation cost while both bounds delimit a neighborhood of \(x_0 \) in catalog size.

%%%
%%% SECTION 4.2
%%%

\subsection{NPE with two summary maps}
\label{subsec:etas-pipelines}

Catalogs simulated from the temporal ETAS model have random length. We compare two ways of compressing a catalog to a fixed-dimensional summary, in terms of the information they retain and the accuracy of the resulting posterior. The first is the hand-crafted map of \citet{stockman2024sb}, which collects \(39\) statistics in three groups. The first five are the log event count \(\ln \{ n(x) \} \), the \(20\)th, \(50\)th and \(90\)th percentiles of the inter-event times \( t_{i+1} - t_i\), and their mean-to-median ratio. The next 18 evaluate Ripley's \(K\) statistic, viz.
\begin{equation}
\label{eq:ripley-k}
\hat K(w) = \frac{T}{n(x)^2}\sum_{i=1}^{n(x)}\sum_{j\neq i}  \mathbf{1} ( 0<t_j-t_i\le w ),
\end{equation}
at windows \(w\) spaced logarithmically below one and linearly above, viz.
\[
w\in\{0.01,0.02,0.04,0.07,0.13,0.22,0.37,0.50,0.72,1,2,4,6,10,15,25,50,100\},
\]
a binning strategy suited to the slow decay of the Omori--Utsu kernel \citep{bacry2016estimation}. The remaining 16 are magnitude-thresholded versions, viz.
\begin{equation}
\label{eq:ripley-kt}
\hat K_T(w,M_T)
=
\frac{T}{\nu^2}\sum_{i:\,m_i\ge M_T}\sum_{j\neq i}
\mathbf{1} ( 0<t_j-t_i\le w ),
\end{equation}
where \(\nu\) is the number of events with magnitude at least \(M_T\). These capture the dependence of clustering on magnitude through the productivity law \eqref{eq:etas-utsu}, and are evaluated on the grid \(M_T\in\{4.0,4.5,5.0,5.5\}\) by \(w\in\{0.2,0.5,1.0,3.0\}\). 

The second compression is learned by an embedding network, which maps a catalog of arbitrary length to a vector of fixed dimension and is trained jointly with the density estimator. The effect of this end-to-end training is discussed in
Section~S1 of the Supplementary Material. The network is built to respect the structure of the data: it processes events in chronological order, weighs them by when they occurred, and handles the varying number of events through a mask, so that padding contributes nothing. Three catalog-level features, namely the event count \(n (X)\), the rate \(n(X)/T\) and \(\ln \{ n (X) \} \), bypass the learned layers and are concatenated to the network's output, so that information about catalog size reaches the density estimator without passing through the compression. The summary again has dimension \(39\), matching the hand-crafted map, so that the two are compared at equal dimension. Further details of the embedding network are given in Section~S3.2 of the Supplementary Material.

Apart from the way a catalog is compressed, the two pipelines are identical. Catalogs are generated by the branching-process algorithm for the ETAS model \citep{zhuang2004,stockman2024sb}, under the prior \eqref{eq:etas-prior} and the restriction \(M\) of Eq.~\eqref{eq:etas-restriction}. The density estimator is a masked autoregressive flow \citep{papamakarios2017masked}, as implemented in the \texttt{sbi} package \citep{tejerocantero2020sbi0}.
The experiment is replicated \(100\) times. In each replication, a parameter \(\theta^\star\) is sampled from the prior, retained only if the branching ratio is below 1 and the expected catalog size lies in \([200,500]\), and a single catalog is simulated from \(p(\cdot\mid\theta^\star)\) to serve as \(x_0\). The restriction \(M\) is thus replication-specific, being defined relative to the corresponding value of \(n(x_0)\). Limiting \(\theta^\star\) to this range keeps the observed catalogs of similar size, so that runtimes are comparable across replications and the experiment can be parallelized under a common time limit. This applies to the choice of \(\theta^\star\) alone; the prior \eqref{eq:etas-prior} is used throughout, both in training and in the MCMC reference.

Within a replication, the estimator is trained over \(20\) rounds, with \(600\) accepted simulations added at each round. Thus, round \(r\) uses a cumulative budget of \(600r\) parameter-data pairs. Parameters are drawn independently from the prior throughout, so the rounds reflect increasing simulation budgets rather than sequential adaptation of the simulation distribution. The training procedure is summarized in Algorithm~1 of the Supplementary Material. The diagnostic of Section~\ref{subsec:gap-diagnostic} is applied at each round to a fresh evaluation set of \(10{,}000\) simulations from the restricted joint, one half supplying the sample from \(P_1\) and the other the sample from \(P_2\). The classifier is a multilayer perceptron with a calibrated probability output, and \(\hat\Delta^M\) is evaluated on draws not used in fitting it.

%%%
%%% SECTION 4.3
%%%

\subsection{Barber--Agakov lower  bounds and the gap diagnostic}
\label{subsec:etas-ba}

\begin{figure}[t!]
\centering
\includegraphics[width=0.95\linewidth]{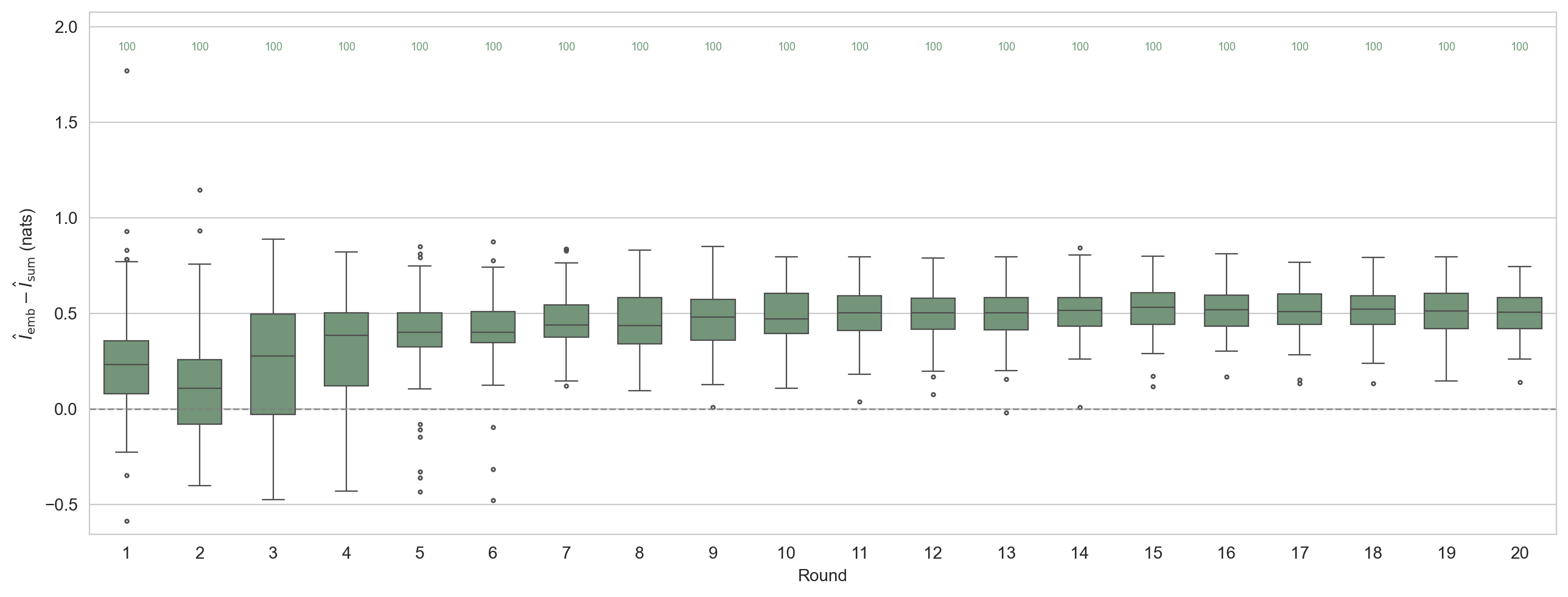}
\caption{Per-replication difference between the two BA lower bounds, \(\hat I(q^M_{\mathrm{emb}},\pi_0^M)-\hat I(q^M_{\mathrm{sum}},\pi_0^M)\), across NPE rounds on \(100\) replications. The dashed line marks zero.}
\label{fig:ba-difference}
\end{figure}

Figure~\ref{fig:ba-difference} reports the per-replication difference between the two BA lower bounds, \(\hat I(q^M_{\mathrm{emb}},\pi_0^M)-\hat I(q^M_{\mathrm{sum}},\pi_0^M)\), across rounds. Only the difference is shown. Each bound separately involves the entropy \(h(\pi_0^M)\) of the restricted prior, which has no closed form and cannot be evaluated by sampling; it cancels between the two, leaving the first bracket of Eq.~\eqref{eq:restricted-gap-difference}. The difference is small and variable over the first few rounds, then rises and settles near \(0.5\) nats. At round 20, its median is \(0.507\) nats with an interquartile range of \([0.420,0.584]\), and it is positive in every replication: on each of the \(100\) replications, the BA lower bound for the learned embedding is the larger.

%%%
%%% FIGURE 1
%%%

\begin{figure}[b!]
\centering
\includegraphics[width=0.95\linewidth]{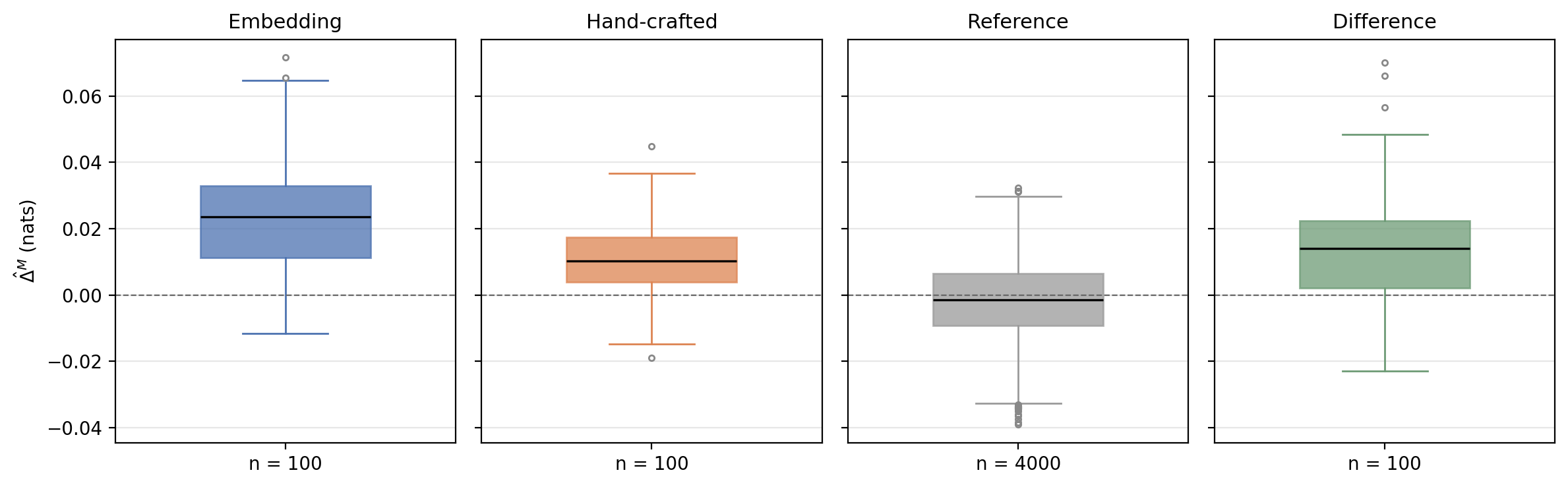}
\caption{Per-replication estimates of the gap \(\Delta^M(s,\phi)\) at round 20, on \(100\) replications: the learned embedding, the hand-crafted summaries, a reference for which the divergence is zero by construction, and the paired difference \(\hat\Delta^M_{\mathrm{emb}}-\hat\Delta^M_{\mathrm{sum}}\), for which positive values indicate that the embedding has the larger gap. The reference pools all replications, rounds and pipelines, hence \(4000\) values rather than \(100\). The dashed line marks zero.}
\label{fig:gap}
\end{figure}

Figure~\ref{fig:gap} reports \(\hat\Delta^M\) at round 20 for each pipeline, together with the reference obtained by applying the same procedure to two samples from \(P_1\), for which the divergence is zero by construction. The reference is centered at \(-0.002\) nats with an interquartile range of \([-0.009, 0.007]\), and \(45\%\) of its values are positive. The classifier therefore introduces no systematic offset, and the spread of these values is the resolution of the diagnostic, against which the two gaps are to be read.

%%%
%%% FIGURE 2
%%%

The two gaps have medians of \(0.010\) nats for the hand-crafted summary and \(0.024\) for the learned embedding, both above the reference median of \(-0.002\), with \(83\%\) and \(95\%\) of values positive against \(45\%\), respectively. This evidence suggests that neither gap is zero, so the assumption that they may be ignored does not hold exactly. Reading Figure~\ref{fig:ba-difference} on its own as a comparison of sufficiency would therefore be unsafe.

What the comparison turns on is the difference between the two gaps, which has median \(0.014\) nats and is positive in \(79\%\) of replications: the summary attaining the larger bound also tends to have the larger gap. By Eq.~\eqref{eq:restricted-gap-difference} the difference between the bounds therefore understates the difference in mutual information, and the comparison of Figure~\ref{fig:ba-difference} is conservative. Had the sign been reversed, part of the apparent advantage would have been attributable to a tighter bound rather than to a more informative summary; the diagnostic is what distinguishes the two cases.

%%%
%%% SECTION 4.4
%%%

\subsection{Consequences for posterior accuracy}
\label{subsec:etas-c2st}

\begin{figure}[b!]
\centering
\includegraphics[width=0.95\linewidth]{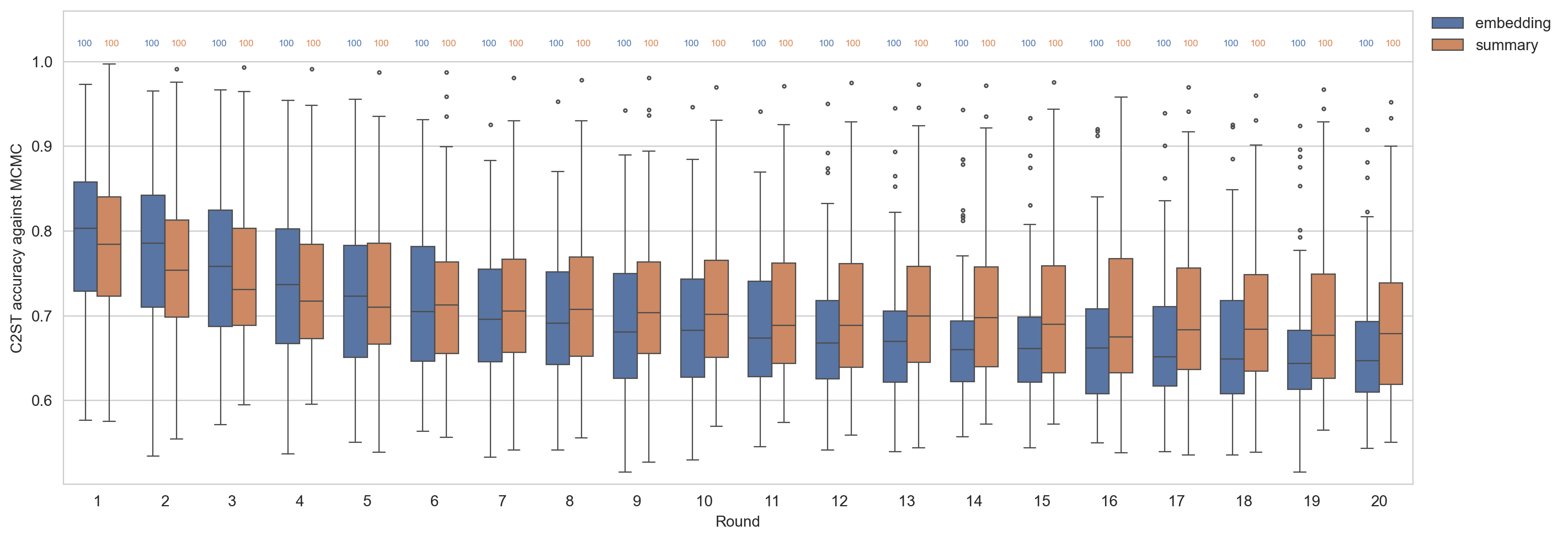}
\caption{Classifier two-sample accuracy between each NPE posterior and the MCMC reference at \(x_0\), across NPE rounds on \(100\) replications.}
\label{fig:c2st}
\end{figure}

The accuracy of the estimated posterior is commonly checked by a classifier two-sample test \citep{lopezpaz2017revisiting}: a classifier is trained to distinguish samples from the NPE posterior at \(x_0\) from samples of the MCMC reference, and its accuracy on held-out draws is reported. This is Eq.~\eqref{eq:tv-identity} applied to a different pair of distributions, so the accuracy again estimates a total variation distance: a value of \(0.5\) means that the two are indistinguishable, and the closer a value is to 0.5, the stronger the agreement. Figure~\ref{fig:c2st} reports this for each pipeline across rounds.

%%%
%%% FIGURE 3
%%%

Both estimated posteriors improve steadily with simulation budget. Over the first five rounds, the posterior conditioned on the hand-crafted summary is narrowly the more accurate, which is what one expects: the embedding network has not yet seen enough simulations to learn a representation that would be competitive with the one designed from domain knowledge. The order reverses at round six and holds for subsequent rounds; at round 20, the posterior conditioned on the learned summary attains a median accuracy of \(0.647\), against \(0.679\) for the hand-crafted one.

The two estimators do not target the same posterior; each approximates the posterior conditioned on its own summary. Figure~\ref{fig:c2st} measures the distance of each to the same full-data posterior \(\pi ( \cdot \mid x_0)\). The gap diagnostic already tells us how well each has learned its own target, and on that count, the estimator built on the hand-crafted summary is the better fitted, its gap being the smaller of the two by a median of \(0.014\) nats. Yet it is the estimated posterior conditioned on the learned summary that is closer to the MCMC reference. The two run in opposite directions, so the advantage of the learned summary cannot be explained by better fitting; since everything else is shared, it must be a consequence of its target sitting closer to \(\pi ( \cdot \mid x_0)\), i.e., from the learned summary discarding less information from the catalog. Therefore, the information-theoretic comparison of Figure~\ref{fig:ba-difference} and the posterior comparison agree, having used no common evidence.

%%%
%%% SECTION 5
%%%

\section{Discussion and Conclusion}
\label{sec:discussion}

In simulation-based inference, when the simulator output is high-dimensional or of varying length, it is standard to compress it into a summary of fixed dimension, and how much information that summary retains directly affects how accurate the resulting posterior estimator can be. MI measures the sufficiency of a summary, but is not computable. It decomposes, however, into a Barber--Agakov lower bound and a non-negative gap, the first of which comes for free with a fitted neural posterior estimator while the second involves the unavailable summary-conditioned posterior. Comparing summaries by their bounds is therefore tempting but incomplete, and this paper has proposed a way of completing it, estimating the gap by classification. We have also demonstrated that the construction remains valid under simulation restricted to a region of the data space containing the observation, with the posterior distribution unchanged for every retained data set.

Section~\ref{subsec:restricted} motivated the restriction on the data space by the cost of simulation and by the relevance of the regime it retains. A further reason concerns the prior. Once \(\pi_0\) has been specified as part of the model, a constraint imposed on the parameter space afterwards, to make simulation feasible, will change the posterior of interest. For example, \citet{stockman2024sb} set out to verify the accuracy of the posterior conditioned on their proposed summary statistics, and choose to assess it against MCMC samples. To keep simulation feasible over the supercritical region, their procedure confines the prior to the subcritical region, replacing the gamma prior on \(\mu\) by a uniform one and imposing \(K\beta_m<\beta_m-\alpha\). The MCMC samples are subject to no such constraint: the latent-branching scheme of \citet{ross2021} evaluates the likelihood of the observed catalog and never simulates, so it is run under the prior as originally specified. The comparison is therefore not entirely fair, the two sets of samples being drawn from posteriors that differ before any summary is introduced and any approximation error is incurred. The data-space restriction avoids this: as shown by Proposition~\ref{prop:posterior-preservation}, the posterior at every retained catalog is unchanged, so the NPE posteriors and the MCMC reference compared in Section~\ref{subsec:etas-c2st} target the same posterior \(\pi(\cdot \mid x_0)\).

The restriction leaves open how \(M\) should be chosen. Apart from requiring that \(M\) contain \(x_0\) and be defined through \(X\) alone, the choice involves a trade-off. A narrower \(M\) concentrates training on catalogs resembling the observation and can improve the fit in that region, but more simulations are discarded, and \(I(\pi_s^M,\pi_0^M)\) then describes a smaller region of the data space. The choice of a larger set \(M\) has the opposite effect. The bounds in Eq.~\eqref{eq:etas-restriction} admit catalogs between one quarter and four times the observed catalog size, wide enough to retain a reasonably broad range of catalogs while still limiting the cost of a supercritical draw. A more systematic choice would be to specify a target acceptance rate and choose an observation-centered region \(M\) so that a prescribed fraction of the simulations is retained. We have not pursued this here, and how best to define such a region is likely to remain problem-specific.

Whatever the choice of \(M\), the diagnostic remains an average. The gap \(\Delta^M(s,\phi)\) \rev{of Section~\ref{subsec:restricted}} is an expectation over catalogs drawn under the restricted joint distribution.
Restricting the data space therefore narrows the region over which this average is taken, but does not make the diagnostic local to \(x_0\). By contrast, the classifier two-sample accuracies of Section~\ref{subsec:etas-c2st} are computed at \(x_0\) alone. Thus the two comparisons in this paper answer different questions: one of them measures retained information on average over the restricted region, while the other compares the two estimated posteriors at the observed catalog. Their agreement is reassuring, but a fully local comparison would require a measure of summary sufficiency defined at \(x_0\). Local posterior-approximation diagnostics that do not require samples from the true posterior are possible, as shown by L-C2ST \citep{linhart2023lc2st}. Developing an analogous local diagnostic for summary sufficiency, defined conditionally at \(x_0\), is left for future work.

Our results support a relative comparison between the two summaries rather than an absolute assessment of sufficiency. They indicate which summary retains more parameter information, but do not determine how much information either has lost relative to the full data. Likewise, the C2ST results show that the learned-summary posterior is closer to the MCMC reference, although neither approximation is indistinguishable from it at the simulation budget considered. Additional diagnostics for the MCMC reference and further analysis of these discrepancies are provided in Section~S4 of the Supplementary Material. More broadly, the diagnostic could be used iteratively to refine summary statistics. Pursuing this for the temporal ETAS model is left for future work. Although NPE is used throughout to construct and compare the summaries, the selected summary can be supplied to other SBI methods that condition on summaries.

%%%
%%% ACKNOWLEDGMENT
%%%
\section*{Acknowledgments}

Renjie Peng gratefully acknowledges financial support from the Fonds de re\-cher\-che du Québec -- Natu\-re et tech\-no\-lo\-gies through a Doctoral Research Scholarship. Additional funding in support of this work was provided through Discovery Grants from the Natural Sciences and Engineering Research Council of Canada to Christian Ge\-nest and David A.~Stephens.

%%%
%%% BIBLIOGRAPHY
%%%

\bibliographystyle{ba}
\bibliography{references}

%%%
%%% SUPPLEMENT
%%%
\clearpage
\setcounter{section}{0}
\setcounter{equation}{0}
\setcounter{figure}{0}
\setcounter{table}{0}
\renewcommand{\thesection}{S\arabic{section}}
\renewcommand{\thesubsection}{S\arabic{section}.\arabic{subsection}}
\renewcommand{\theequation}{S\arabic{equation}}
\renewcommand{\thefigure}{S\arabic{figure}}
\renewcommand{\thetable}{S\arabic{table}}
\renewcommand{\theHsection}{S\arabic{section}}
\renewcommand{\theHequation}{S\arabic{equation}}
\renewcommand{\theHfigure}{S\arabic{figure}}
\renewcommand{\theHtable}{S\arabic{table}}

\begin{center}
\textbf{\Large Supplementary Material}
\end{center}

\section{End-to-End Training of Learned Summaries}
\label{sec:supp-e2e}

This section examines what end-to-end training optimizes when the summary is learned rather than fixed. The objective turns out to be the Barber--Agakov (BA) lower bound on the mutual information carried by the embedding, so that a single loss jointly optimizes the summary and the density estimator. The tightness of that bound, however, is no better controlled than in the fixed-summary case.

The notation is that of the main text, specialized to a learned summary by setting \(s = s(\cdot,\eta)\). We write \(\pi_\eta\) for the resulting summary-conditioned posterior, \(q_{\eta,\phi}(\cdot\mid X) = q_{F\{s(X,\eta),\phi\}} \) for its estimator, and \(\hat I(q_{\eta,\phi},\pi_0)\) and \(\Delta(\eta,\phi)\) for the corresponding BA lower bound and gap, respectively, so that
\begin{equation}
\label{eq:supp-ba-bound}
\hat I(q_{\eta,\phi},\pi_0) = \int_{\mathcal{X}}\!\int_{\Theta} \ln q_{\eta,\phi}(\theta\mid x)\, \pi_0(\theta)\,p(x\mid\theta)\,\mathrm{d}\theta\,\mathrm{d}x + h(\pi_0)
\end{equation}
and
\begin{equation}
\label{eq:supp-ba-decomposition}
I(\pi_\eta,\pi_0)=\hat I(q_{\eta,\phi},\pi_0)+\Delta(\eta,\phi),
\end{equation}
with \(h(\pi_0)\) the differential entropy of the prior.

Minimizing the negative log-density loss
\[
- \frac{1}{N} \sum_{i=1}^{N}\ln \{ q_{\eta,\phi}(\theta_i\mid X_i) \}
\]
over the simulated pairs corresponds, at the population level, to maximizing
\begin{equation}
\label{eq:supp-joint-objective}
\mathcal{J}(\eta,\phi)
=
\int_{\mathcal{X}}\!\int_{\Theta} \ln \{ q_{\eta,\phi}(\theta\mid x)\} \, \pi_0(\theta)\,p(x\mid\theta)\,\mathrm{d}\theta\,\mathrm{d}x .
\end{equation}
Comparing Eq.~\eqref{eq:supp-joint-objective} with Eq.~\eqref{eq:supp-ba-bound} shows that \(\mathcal{J}(\eta,\phi)=\hat I(q_{\eta,\phi},\pi_0)-h(\pi_0)\), so that by Eq.~\eqref{eq:supp-ba-decomposition},
\begin{equation}
\label{eq:supp-objective-rewritten}
\mathcal{J}(\eta,\phi)
=
I(\pi_\eta,\pi_0)-\Delta(\eta,\phi)-h(\pi_0).
\end{equation}

Since \(h(\pi_0)\) is constant and the mutual information does not depend on \(\phi\), one has 
\[
\nabla_\phi\mathcal{J}=-\,\nabla_\phi\Delta(\eta,\phi),
\]
so updating \(\phi\) reduces the gap at the current embedding, driving the estimator towards \(\pi_\eta\). Both terms depend on \(\eta\), giving
\[
\nabla_\eta\mathcal{J}
=
\nabla_\eta I(\pi_\eta,\pi_0)-\nabla_\eta\Delta(\eta,\phi),
\]
so updating \(\eta\) encourages the embedding to retain information about \(\theta\), but only in a form that the current estimator can represent. Neither term is computed separately in practice; both arise from the single gradient of \(\mathcal{J}\), and the decomposition serves only to interpret it.

Training therefore maximizes \(\hat I(q_{\eta,\phi},\pi_0)\), not \(I(\pi_\eta,\pi_0)\) itself. The two differ by \(\Delta(\eta,\phi)\), so a large value of the objective is no evidence that the embedding is close to sufficient, and assessing the gap requires the diagnostic of Section~3 of the main text.

%%%
%%% SECTION 2
%%%

\section{Classifier-Based Identities for Total Variation and Kullback--Leibler (KL) Divergence}
\label{sec:supp-classifier}

This section proves the two identities stated in the main text. Throughout, the pair \((\theta,s)\) ranges over \(\Theta\times\mathcal{S}\) and
\[
P_1(\theta,s)=\pi(\theta\mid s)\,m_s(s),
\qquad
P_2(\theta,s)=q_{F(s,\phi)}(\theta)\,m_s(s)
\]
are the two joint densities obtained by pairing the summary marginal \(m_s\) with the true and the estimated summary-conditioned posterior. Both distributions can be sampled using the same simulator. For \(P_1\), draw
\[
\theta\sim\pi_0,
\qquad
X\sim p(\cdot\mid\theta),
\]
and form the pair \((\theta,s(X))\). For \(P_2\), generate \(X\) in the same way, discard the corresponding parameter draw, and replace it by
\[
\theta'\sim q_{F(s(X),\phi)}.
\]
Then the pair \((\theta',s(X))\) is a draw from \(P_2\). Throughout, \(P_1\) is assumed to be absolutely continuous with respect to \(P_2\). The quantity to be identified is
the gap
\begin{equation}
\label{eq:supp-gap}
\Delta(s,\phi) = \int_{\mathcal{S}}
D_{\mathrm{KL}}\bigl\{\pi(\cdot\mid s)\,\big\|\,q_{F(s,\phi)}(\cdot)\bigr\}\, m_s(s)\,\mathrm{d}s = D_{\mathrm{KL}}(P_1\,\|\,P_2),
\end{equation}
where the second equality holds because \(P_1\) and \(P_2\) share the marginal \(m_s\)
and differ only in their conditionals.

Let \(Y\) be equally likely to be \(0\) or \(1\), and let the pair \((\theta,s)\) be drawn from \(P_1\) when \(Y=1\) and from \(P_2\) when \(Y=0\). Further, let \(d : \Theta\times\mathcal{S}\to (0,1)\) be trained to predict \(Y\) by minimizing the population binary cross-entropy
\[
L(d)=-\mathbb{E}[Y\ln d(\theta,s)+(1-Y)\ln\{1-d(\theta,s)\}],
\] 
which under equal
class probabilities equals
\begin{equation}
\label{eq:supp-bce}
L(d)
=
-\tfrac12\int_{\Theta\times\mathcal{S}}\ln\{d(\theta,s)\}\,
P_1(\theta,s)\,\mathrm{d}\theta\,\mathrm{d}s
-\tfrac12\int_{\Theta\times\mathcal{S}}\ln\{1-d(\theta,s)\}\,
P_2(\theta,s)\,\mathrm{d}\theta\,\mathrm{d}s .
\end{equation}

\paragraph{The Bayes-optimal classifier.}
The integrand of Eq.~\eqref{eq:supp-bce} is convex in \(d(\theta,s)\) on \((0,1)\)
and can be minimized pointwise, at the stationary point
\[
-\frac{1}{2} \,\frac{P_1(\theta,s)}{d(\theta,s)}
+ \frac{1}{2} \,\frac{P_2(\theta,s)}{1-d(\theta,s)}=0 .
\]
Solving this equation yields
\begin{equation}
\label{eq:supp-bayes-classifier}
d^\star(\theta,s)=\frac{P_1(\theta,s)}{P_1(\theta,s)+P_2(\theta,s)},
\quad\text{or equivalently}\quad
\frac{d^\star(\theta,s)}{1-d^\star(\theta,s)}=\frac{P_1(\theta,s)}{P_2(\theta,s)} .
\end{equation}
The same \(d^\star\) underlies both identities below; they differ only in how its output is post-processed.

\paragraph{Total variation from the decision rule.}

The Bayes-optimal decision rule is \(\hat Y=\mathbf{1} (d^\star>1/2) = \mathbf{1} (P_1>P_2) \), i.e., predict whichever class has higher density at \((\theta,s)\). Under equal class probabilities, the marginal density of \((\theta,s)\) is \(\bar m=(P_1+P_2)/2\), and Bayes' rule gives the class posteriors
\[
\Pr(Y = 1\mid\theta,s) = \frac{P_1}{P_1+P_2},
\qquad
\Pr(Y=0\mid\theta,s)=\frac{P_2}{P_1+P_2} .
\]
The rule assigns each point to the class with the larger posterior, so the probability of a correct prediction at \((\theta,s)\) is the larger of the two, i.e., \(\max(P_1,P_2)/(P_1+P_2)\). Marginalizing and using the elementary identity
\(\max(a,b) = (a+b)/2 + |a-b|/2\), one finds
\[
\begin{split}
\Pr(Y = \hat Y)
&=
\int_{\Theta\times\mathcal{S}}\bar m\,\frac{\max(P_1,P_2)}{P_1+P_2}
\,\mathrm{d}\theta\,\mathrm{d}s
=
\frac{1}{2} \int_{\Theta\times\mathcal{S}}\max(P_1,P_2)\,\mathrm{d}\theta\,\mathrm{d}s\\[2pt]
&=
\frac{1}{4} \int_{\Theta\times\mathcal{S}}(P_1+P_2)\,\mathrm{d}\theta\,\mathrm{d}s
+ \frac{1}{4} \int_{\Theta\times\mathcal{S}}|P_1-P_2|\,\mathrm{d}\theta\,\mathrm{d}s
=
\frac{1}{2} + \frac{1}{2} \, \mathrm{TV}(P_1,P_2),
\end{split}
\]
where the last step uses \(\int(P_1+P_2)=2\) and the \(L^1\) definition
\[
\mathrm{TV}(P_1,P_2) = \frac{1}{2} \int_{\Theta\times\mathcal{S}}|P_1-P_2|.
\]
This is the first identity.

\paragraph{Divergence from the log-ratio.}
Taking logarithms in the second equality of Eq.~\eqref{eq:supp-bayes-classifier} and
integrating against \(P_1\),
\begin{align*}
\int_{\Theta\times\mathcal{S}}
\ln\!\left\{\frac{d^\star(\theta,s)}{1-d^\star(\theta,s)}\right\}
P_1(\theta,s)\,\mathrm{d}\theta\,\mathrm{d}s
& =
\int_{\Theta\times\mathcal{S}}
\ln\!\left\{\frac{P_1(\theta,s)}{P_2(\theta,s)}\right\}
P_1(\theta,s)\,\mathrm{d}\theta\,\mathrm{d}s
\\[3mm]
& =
D_{\mathrm{KL}}(P_1\,\|\,P_2)
=
\Delta(s,\phi)
\end{align*}
by Eq.~\eqref{eq:supp-gap}. This is the second identity. \(\square\)

\paragraph{Plug-in estimators.}

In practice, \(d^\star\) is unknown and is replaced by the classifier \(\hat d\) obtained by minimizing the empirical version of Eq.~\eqref{eq:supp-bce}. Given a random sample $(\theta_1, s_1), \dots, (\theta_n, s_n)$ from $P_1$, one can set
\[
\hat\Delta = \frac{1}{n}\sum_{i=1}^{n} \ln\!\left\{ \frac{\hat d(\theta_i,s_i)} {1-\hat d(\theta_i,s_i)} \right\}.
\]
The average is computed over held-out draws not used to fit \(\hat d\). 

In this study, the classifier used for the BA-gap diagnostic is a two-hidden-layer multilayer perceptron with \(64\) units per layer and ReLU activations, followed by five-fold isotonic probability calibration.

%%%
%%% SECTION 3
%%%

\section{Implementation Details for the Temporal ETAS Experiment}
\label{sec:supp-implementation}

\subsection{Cumulative-budget amortized Neural Posterior Estimation (NPE) under data-space restriction}
\label{subsec:supp-cumulative-npe}

In the temporal ETAS experiment, the term ``round'' indexes increasing cumulative simulation budgets. Parameters are drawn from the original prior \(\pi_0\) throughout, and data are generated from the same simulator \(p(\cdot\mid\theta)\) at every round. At each round, newly accepted simulations are added to those accumulated previously.
Algorithm~\ref{alg:supp-cumulative-npe} summarizes the procedure.

\begin{algorithm}[h!]
\caption{Cumulative-budget amortized NPE under data-space restriction}
\label{alg:supp-cumulative-npe}
\begin{algorithmic}[1]

\Require Prior \(\pi_0\), simulator \(p(\cdot\mid\theta)\), data-space region \(M\),
summary map \(s\), number of rounds \(R\), and number of accepted simulations per round \(N\)

\State \(\mathcal{D}_0 \gets \varnothing\)

\For{\(r \in \{ 1, \dots, R\} \)}

    \State \(\mathcal{D}^{\mathrm{new}}_r \gets \varnothing\)

    \While{\(|\mathcal{D}^{\mathrm{new}}_r|<N\)}
        \State Sample \(\theta\sim\pi_0\)
        \State Simulate \(X\sim p(\cdot\mid\theta)\)
        \If{\(X\in M\)}
            \State Add \((\theta,X)\) to \(\mathcal{D}^{\mathrm{new}}_r\)
        \EndIf
    \EndWhile

    \State \(\mathcal{D}_r
    \gets
    \mathcal{D}_{r-1}\cup\mathcal{D}^{\mathrm{new}}_r\)

    \State Train \(q^M_{s,\phi}\) on \(\mathcal{D}_r\) by minimizing
    \[
    \mathcal{L}_r^M(\phi)
    =
    -\frac{1}{rN}
    \sum_{k=1}^{r}\sum_{j=1}^{N}
    \ln \{ q^M_{s,\phi}(\theta_{k,j}\mid X_{k,j})\}.
    \]

\EndFor

\end{algorithmic}
\end{algorithm}

\subsection{Architecture of the embedding network}

\label{subsec:supp-architecture}
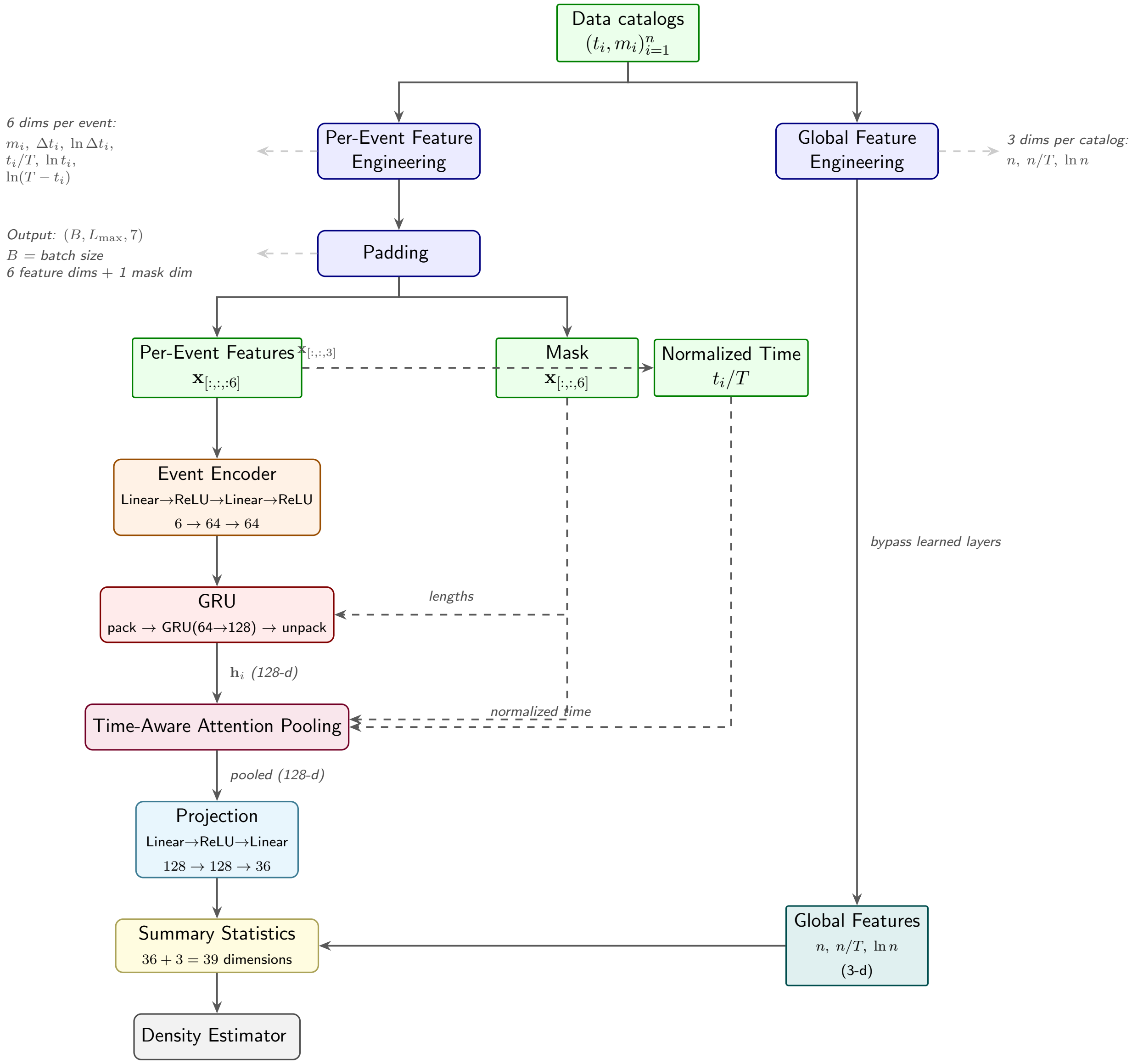
\begin{figure}[t!]
\centering
\begin{tikzpicture}[
scale=0.73, transform shape,
block/.style={
rectangle, draw, rounded corners=3pt,
minimum width=3.2cm, minimum height=0.9cm,
align=center, font=\small\sffamily,
fill=blue!8, draw=blue!50!black, line width=0.6pt
},
data/.style={
rectangle, draw, rounded corners=1pt,
minimum width=2.8cm, minimum height=0.7cm,
align=center, font=\small\sffamily,
fill=green!8, draw=green!50!black, line width=0.6pt
},
globaldata/.style={
rectangle, draw, rounded corners=1pt,
minimum width=2.8cm, minimum height=0.7cm,
align=center, font=\small\sffamily,
fill=teal!12, draw=teal!60!black, line width=0.6pt
},
arrow/.style={
-{Stealth[length=5pt, width=4pt]},
line width=0.7pt, draw=gray!70!black
},
dasharrow/.style={
-{Stealth[length=5pt, width=4pt]},
line width=0.7pt, draw=gray!70!black, dashed
},
label/.style={
font=\scriptsize\sffamily\itshape, text=gray!60!black
},
every node/.append style={font=\small\sffamily}
]

\node[data] (catalogs) {
Data catalogs\\
$(t_i,m_i)_{i=1}^{n}$
};

\node[block, below left=1.2cm and 1.5cm of catalogs]
(feateng_event) {
Per-Event Feature\\
Engineering
};

\node[block, below right=1.2cm and 1.5cm of catalogs]
(feateng_global) {
Global Feature\\
Engineering
};

\draw[arrow] (catalogs.south) -- ++(0,-0.4) -| (feateng_event.north);
\draw[arrow] (catalogs.south) -- ++(0,-0.4) -| (feateng_global.north);

\node[label, left=1.2cm of feateng_event,
text width=4.8cm, align=left]
(event_feat_list) {%
6 dims per event:\\[2pt]
$m_i,\;\Delta t_i,\;\ln\Delta t_i,$\\
$t_i/T,\;\ln t_i,$\\
$\ln(T-t_i)$%
};

\draw[dasharrow, gray!40]
(feateng_event.west) -- (event_feat_list.east);

\node[label, right=1.2cm of feateng_global,
text width=3.8cm, align=left]
(global_feat_list) {%
3 dims per catalog:\\[2pt]
$n,\;n/T,\;\ln n$%
};

\draw[dasharrow, gray!40]
(feateng_global.east) -- (global_feat_list.west);

\node[block, below=1.0cm of feateng_event]
(padding) {
Padding
};

\draw[arrow] (feateng_event) -- (padding);

\node[label, left=1.2cm of padding,
text width=4.8cm, align=left]
(pad_annot) {%
Output: $(B,L_{\max},7)$\\[2pt]
$B$ = batch size\\
6 feature dims $+$ 1 mask dim%
};

\draw[dasharrow, gray!40]
(padding.west) -- (pad_annot.east);

\node[data, below left=1.2cm and 0.3cm of padding]
(features) {
Per-Event Features\\
$\mathbf{x}_{[:,:,:6]}$
};

\node[data, below right=1.2cm and 0.3cm of padding]
(mask) {
Mask\\
$\mathbf{x}_{[:,:,6]}$
};

\draw[arrow] (padding.south) -- ++(0,-0.4) -| (features.north);
\draw[arrow] (padding.south) -- ++(0,-0.4) -| (mask.north);

\node[data, right=0.3cm of mask]
(reltime) {
Normalized Time\\
$t_i/T$
};

\draw[dasharrow]
(features.east) -- ++(0.3,0) |- (reltime.west)
node[label, pos=0.25, above=1pt]
{$\mathbf{x}_{[:,:,3]}$};

\node[
block,
below=1.2cm of features,
fill=orange!10,
draw=orange!60!black
]
(encoder) {%
Event Encoder\\
{\scriptsize Linear$\to$ReLU$\to$Linear$\to$ReLU}\\
{\scriptsize $6\to64\to64$}%
};

\draw[arrow] (features) -- (encoder);

\node[
block,
below=1.0cm of encoder,
fill=red!8,
draw=red!50!black
]
(gru) {%
GRU\\
{\scriptsize pack $\to$ GRU(64$\to$128) $\to$ unpack}%
};

\draw[arrow] (encoder) -- (gru);

\draw[dasharrow]
(mask.south) |- (gru.east)
node[label, near end, above=2pt]
{\scriptsize lengths};

\node[
block,
below=1.2cm of gru,
fill=purple!10,
draw=purple!60!black,
minimum width=4.5cm,
minimum height=0.9cm
]
(tap) {%
Time-Aware Attention Pooling%
};

\draw[arrow]
(gru) -- (tap)
node[label, midway, right=3pt]
{$\mathbf{h}_i$ (128-d)};

\draw[dasharrow]
(reltime.south) |- (tap.east)
node[label, near end, above=2pt]
{\scriptsize normalized time};

\draw[dasharrow]
(mask.south) |- ([yshift=0.15cm]tap.east);

\node[
block,
below=1.0cm of tap,
fill=cyan!8,
draw=cyan!50!black
]
(proj) {%
Projection\\
{\scriptsize Linear$\to$ReLU$\to$Linear}\\
{\scriptsize $128\to128\to36$}%
};

\draw[arrow]
(tap) -- (proj)
node[label, midway, right=3pt]
{pooled (128-d)};

\node[
block,
below=0.8cm of proj,
fill=yellow!15,
draw=yellow!60!black,
minimum width=4.0cm
]
(summary) {%
Summary Statistics\\
{\scriptsize $36+3=39$ dimensions}%
};

\draw[arrow]
(proj) -- (summary)
node[label, midway, left=3pt]
{ };

\node[globaldata]
(global_vec) at (feateng_global |- summary) {%
Global Features\\
{\scriptsize $n,\;n/T,\;\ln n$}\\
{\scriptsize (3-d)}%
};

\draw[arrow]
(feateng_global.south) -- (global_vec.north)
node[label, midway, right=3pt]
{\scriptsize bypass learned layers};

\draw[arrow]
(global_vec) -- (summary)
node[label, midway, above=2pt]
{};

\node[
block,
below=0.8cm of summary,
fill=gray!10,
draw=gray!60!black
]
(maf) {
Density Estimator
};

\draw[arrow] (summary) -- (maf);

\end{tikzpicture}

\caption{Architecture of the embedding network. Per-event features are padded to
\(L_{\max}=4n(x_0)\) and passed through the learned layers, while three
catalog-level features bypass these layers and are concatenated with the
\(36\)-dimensional learned representation to form the \(39\)-dimensional
summary. Solid arrows indicate the main data flow; dashed arrows indicate
secondary inputs.}

\label{fig:embedding-architecture}
\end{figure}

Figure~\ref{fig:embedding-architecture} shows the embedding network used for the
learned summary of Section~4.2 of the main text. A catalog is passed through two
branches. The per-event branch computes six features for each event, pads the
sequence to a common length with a trailing mask channel, and passes the result
through an encoder, a gated recurrent unit (GRU) and a time-aware attention pooling layer
before projecting to \(36\) dimensions. The global branch computes three
catalog-level features, which bypass all learned layers and are concatenated
directly, giving a summary of dimension \(39\). The mask enters both the recurrent
layer, where it supplies sequence lengths, and the pooling layer, where it prevents
padded positions from contributing to the attention weights; the normalized time
\(t_i/T\) enters the pooling layer as an additional score input, so
that the weight given to an event may depend on when it occurred. The network is
trained jointly with the density estimator by the procedure of
Section~\ref{sec:supp-e2e}.

%%%
%%% Figure S1
%%%
\vspace{2em}

% \clearpage

%%%
%%% SECTION 4
%%%

\section{Diagnosing the C2ST Discrepancy}
\label{sec:supp-c2st-diagnosis}

\subsection{Diagnostics for the Markov chain Monte Carlo (MCMC) reference}
\label{subsec:supp-mcmc-diagnostics}

Because the C2ST accuracies remain separated from their chance level of \(0.5\), we first assess whether the MCMC output provides a reliable reference for this comparison. The sampler uses the exact ETAS likelihood and the same prior as the NPE analysis. For each of the \(100\) replications, we ran six chains, with one initialized at the maximum-likelihood estimate and the remaining five initialized at dispersed points in the prior support. Each chain provided \(5000\) post-warmup draws. Across all replications and parameters, the largest rank-normalized split-\(\widehat R\) was \(1.018\). Six replications had at least one parameter with \(\widehat R>1.01\), but none had \(\widehat R>1.05\), providing no indication of substantial disagreement among the chains.

Agreement among chains does not by itself guarantee sufficient Monte Carlo precision. We therefore also computed the bulk and tail effective sample sizes for every parameter. The smallest bulk and tail effective sample sizes across all replications were \(494\) and \(901\), respectively, and neither diagnostic fell below \(400\). Thus, even for the most strongly autocorrelated parameter chains, the retained output contains several hundred effective draws for estimating the center and tails of the posterior.

Finally, we examined whether the few replications with \(\widehat R>1.01\) coincided with those showing the largest discrepancy from the MCMC reference. None of the six replications with \(\widehat R>1.01\) belonged to the ten replications with the largest round-20 C2ST accuracy for either NPE pipeline. The largest C2ST discrepancies therefore do not occur where the MCMC mixing diagnostics are weakest. Taken together, the chain agreement, effective sample sizes, and lack of overlap with the worst-C2ST replications provide no evidence that inadequate MCMC mixing explains the departure of the C2ST accuracy from \(0.5\).

\subsection{Characterizing the discrepancies}

To investigate the remaining discrepancies between the NPE approximations and the MCMC reference, we focus on the final training round and examine how these discrepancies vary across replications for both pipelines. We first consider whether larger discrepancies are associated with particular regions of the ETAS parameter space, and then compare posterior means, standard deviations, and pairwise correlations with the MCMC reference to assess whether the discrepancies are primarily reflected in posterior location, scale, or dependence structure. In addition, the posterior C2ST comparisons use logistic regression as the classifier.

Tables~S1--S2 show that the largest round-20 C2ST discrepancies occur across a broad range of ETAS parameter values rather than being confined to a single parameter regime. The sets of worst-performing replications also overlap only partially between the two pipelines, suggesting that the sources of approximation error are not identical for the embedding and summary approaches.

%%%
%%% TABLE S1
%%%

\begin{table}[H]
\centering
\caption{The ten replications with the largest round-20 C2ST accuracy for the embedding pipeline, ordered from largest down. Columns give the true parameters, the branching ratio $n^\ast$, the number of observed events $n$, and the round-20 C2ST accuracy of the embedding (Emb.) and summary (Sum.) pipelines on the same replication. The last three rows give the minimum, median, and maximum over the embedding pipeline's 25 most accurate replications.\\[2mm]}
\label{tab:supp-worst10-emb}
\small
\setlength{\tabcolsep}{4pt}
\begin{tabular}{lrrrrrrrrr}
\toprule
 & & & & & & & & \multicolumn{2}{c}{Round-20 C2ST} \\
\cmidrule(lr){9-10}
Seed & $\mu$ & $K$ & $\alpha$ & $c$ & $p$ & $n^\ast$ & $n$ & Emb. & Sum. \\
\midrule
89 & 0.074 & 0.387 & 1.231 & 0.0059 & 1.563 & 0.83 & 448 & 0.920 & 0.900 \\
84 & 0.212 & 0.328 & 0.783 & 0.0177 & 1.848 & 0.50 & 428 & 0.882 & 0.802 \\
94 & 0.050 & 0.396 & 1.222 & 0.0071 & 1.363 & 0.84 & 340 & 0.863 & 0.890 \\
41 & 0.129 & 0.261 & 1.102 & 0.0202 & 1.641 & 0.50 & 246 & 0.822 & 0.860 \\
70 & 0.297 & 0.096 & 1.139 & 0.0315 & 1.698 & 0.19 & 386 & 0.817 & 0.952 \\
88 & 0.057 & 0.355 & 1.189 & 0.0360 & 1.679 & 0.73 & 244 & 0.814 & 0.728 \\
64 & 0.162 & 0.119 & 0.963 & 0.0388 & 1.855 & 0.20 & 233 & 0.756 & 0.638 \\
78 & 0.071 & 0.458 & 1.018 & 0.0427 & 1.404 & 0.82 & 299 & 0.748 & 0.770 \\
24 & 0.148 & 0.438 & 0.586 & 0.0384 & 1.829 & 0.59 & 381 & 0.744 & 0.685 \\
98 & 0.267 & 0.150 & 1.387 & 0.0199 & 1.901 & 0.38 & 341 & 0.734 & 0.736 \\
\midrule
Best-25 min & 0.050 & 0.012 & 0.537 & 0.0066 & 1.050 & 0.02 & 68 & 0.544 & 0.551 \\
Best-25 median & 0.176 & 0.143 & 1.043 & 0.0253 & 1.338 & 0.23 & 276 & 0.592 & 0.648 \\
Best-25 max & 0.478 & 0.384 & 1.487 & 0.0487 & 1.935 & 0.86 & 521 & 0.610 & 0.822 \\
\bottomrule
\end{tabular}
\end{table}

\clearpage

Figures~S2--S3 provide a graphical view of the same comparison in the ETAS parameter space, showing where the worst-performing and most accurate replications lie relative to the remaining cases and whether they cluster in particular parameter regimes.

%%%
%%% TABLE S2
%%%

\begin{table}[t!]
\centering
\caption{As Table~\ref{tab:supp-worst10-emb}, for the summary pipeline: the ten replications with its largest round-20 C2ST accuracy, and the minimum, median, and maximum over its twenty-five most accurate replications.\\[2mm]}
\label{tab:supp-worst10-sum}
\small
\setlength{\tabcolsep}{4pt}
\begin{tabular}{lrrrrrrrrr}
\toprule
 & & & & & & & & \multicolumn{2}{c}{Round-20 C2ST} \\
\cmidrule(lr){9-10}
Seed & $\mu$ & $K$ & $\alpha$ & $c$ & $p$ & $n^\ast$ & $n$ & Emb. & Sum. \\
\midrule
70 & 0.297 & 0.096 & 1.139 & 0.0315 & 1.698 & 0.19 & 386 & 0.817 & 0.952 \\
68 & 0.133 & 0.285 & 1.356 & 0.0408 & 1.768 & 0.69 & 351 & 0.685 & 0.933 \\
89 & 0.074 & 0.387 & 1.231 & 0.0059 & 1.563 & 0.83 & 448 & 0.920 & 0.900 \\
94 & 0.050 & 0.396 & 1.222 & 0.0071 & 1.363 & 0.84 & 340 & 0.863 & 0.890 \\
100 & 0.253 & 0.302 & 0.789 & 0.0069 & 1.975 & 0.46 & 478 & 0.664 & 0.862 \\
41 & 0.129 & 0.261 & 1.102 & 0.0202 & 1.641 & 0.50 & 246 & 0.822 & 0.860 \\
14 & 0.073 & 0.295 & 1.498 & 0.0454 & 1.187 & 0.84 & 252 & 0.726 & 0.846 \\
79 & 0.212 & 0.045 & 1.423 & 0.0066 & 1.133 & 0.12 & 282 & 0.582 & 0.822 \\
71 & 0.101 & 0.277 & 1.424 & 0.0176 & 1.202 & 0.73 & 206 & 0.682 & 0.820 \\
66 & 0.410 & 0.060 & 0.938 & 0.0257 & 1.913 & 0.10 & 497 & 0.664 & 0.813 \\
\midrule
Best-25 min & 0.050 & 0.020 & 0.549 & 0.0087 & 1.050 & 0.04 & 68 & 0.544 & 0.551 \\
Best-25 median & 0.175 & 0.205 & 0.781 & 0.0283 & 1.338 & 0.32 & 260 & 0.638 & 0.595 \\
Best-25 max & 0.349 & 0.477 & 1.435 & 0.0493 & 1.971 & 0.86 & 400 & 0.718 & 0.617 \\
\bottomrule
\end{tabular}
\end{table}

Figures~S4--S5 show that posterior means generally track the MCMC reference, with larger departures occurring for some replications and parameters. Figures~S6--S7 reveal more systematic differences in posterior scale, with the NPE posteriors often wider than the MCMC reference, particularly for the summary pipeline. Across both comparisons, the most noticeable departures are seen for \(\alpha\) in the summary pipeline. This may be related to the hand-crafted treatment of magnitude information: the magnitude-threshold versions of Ripley's \(K\) reduce magnitudes to exceedances of a small number of fixed thresholds, potentially discarding continuous information that is particularly relevant for \(\alpha\). Figures~S8--S9 show that pairwise posterior correlations broadly follow the MCMC reference, although the worst-performing replications appear somewhat more concentrated for correlations between \(c\) and \(p\).

%%%
%%% FIGURE S2
%%%

\begin{figure}[b!]
\centering
\includegraphics[width=0.45\linewidth]{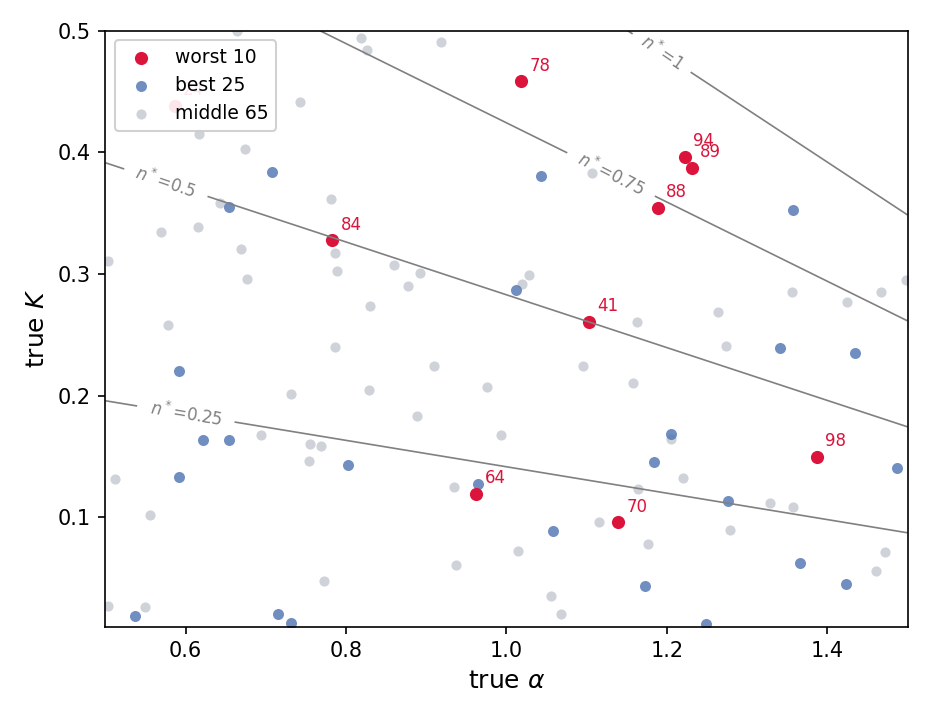}\hfill \includegraphics[width=0.45\linewidth]{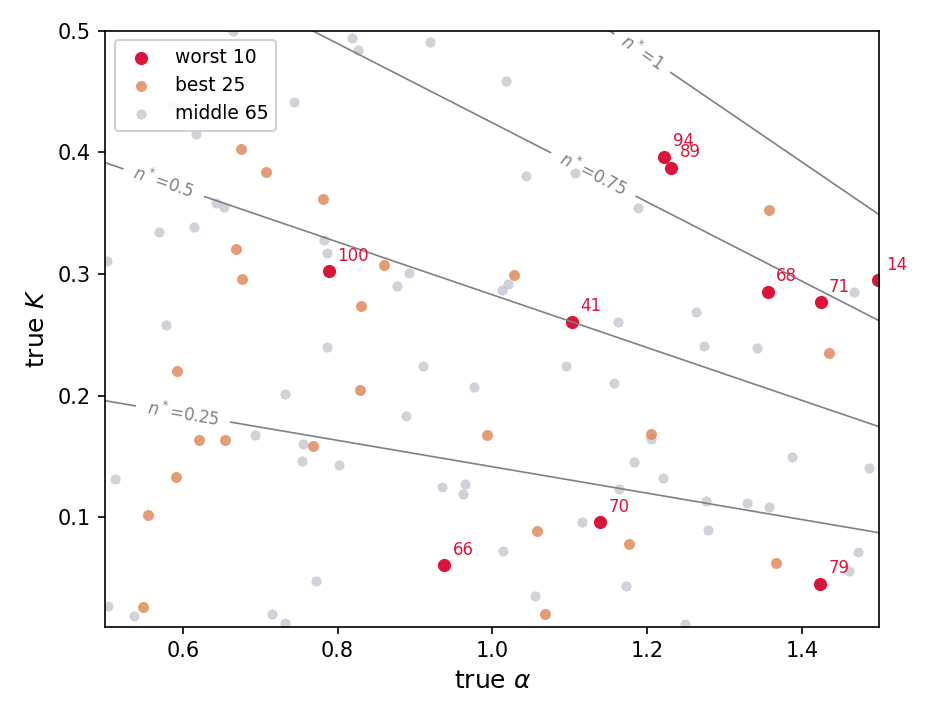} \caption{True \((\alpha,K)\) of each replication, with contours of the branching ratio \(n^\ast=K\beta_m/(\beta_m-\alpha)\). Red: the 10 replications with the largest round-20 C2ST accuracy of the embedding (left) and summary (right) pipeline, labelled by seed; blue (embedding) and orange (summary): the 25 most accurate for that pipeline; grey: the remaining 65.}
\label{fig:supp-Kalpha}
\end{figure}

%%%
%%% FIGURE S3
%%%

\begin{figure}[t!]
\centering
\includegraphics[width=0.45\linewidth]{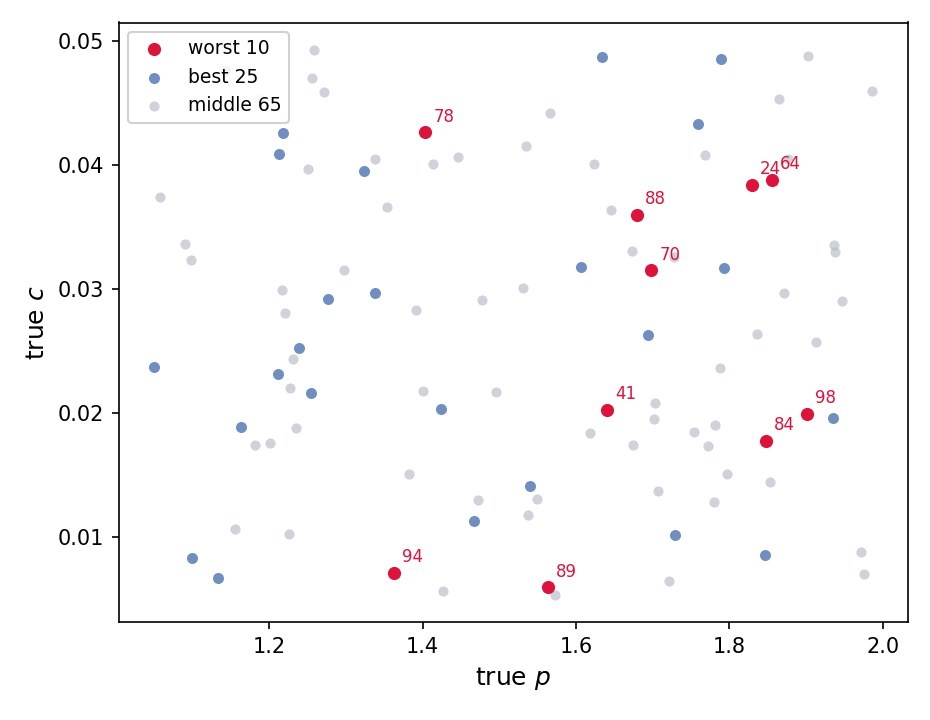}\hfill
\includegraphics[width=0.45\linewidth]{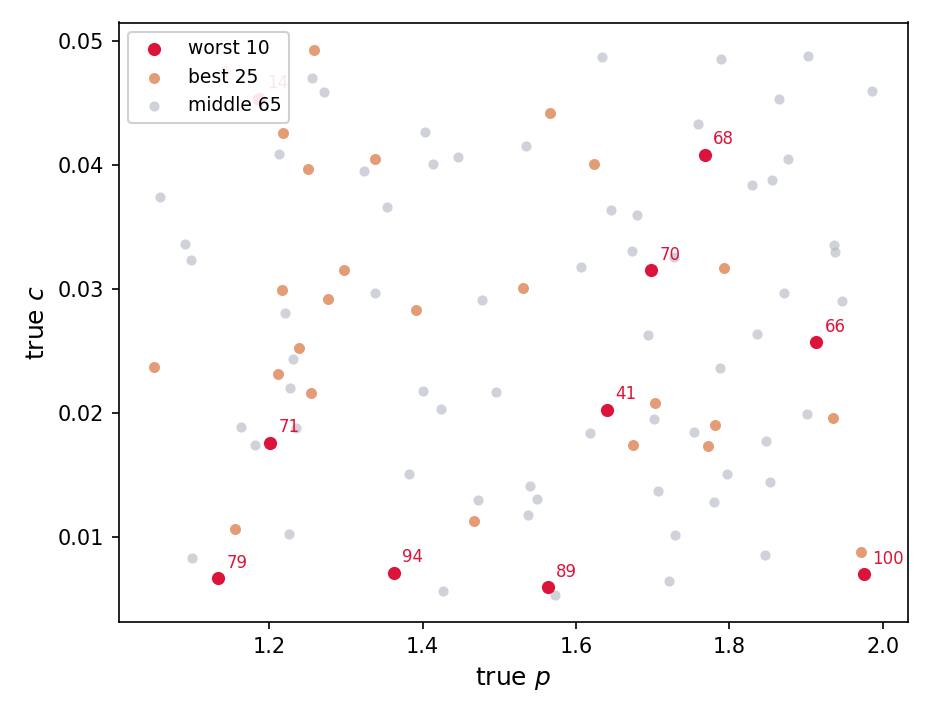}
\caption{As Figure~\ref{fig:supp-Kalpha}, in the \((p,c)\) plane. Small \(c\) and large \(p\) correspond to aftershocks arriving tightly packed behind their parent event.}
\label{fig:supp-cp}
\end{figure}

%%%
%%% FIGURE S4
%%%

\begin{figure}[b!]
\centering
\includegraphics[width=0.85\linewidth]{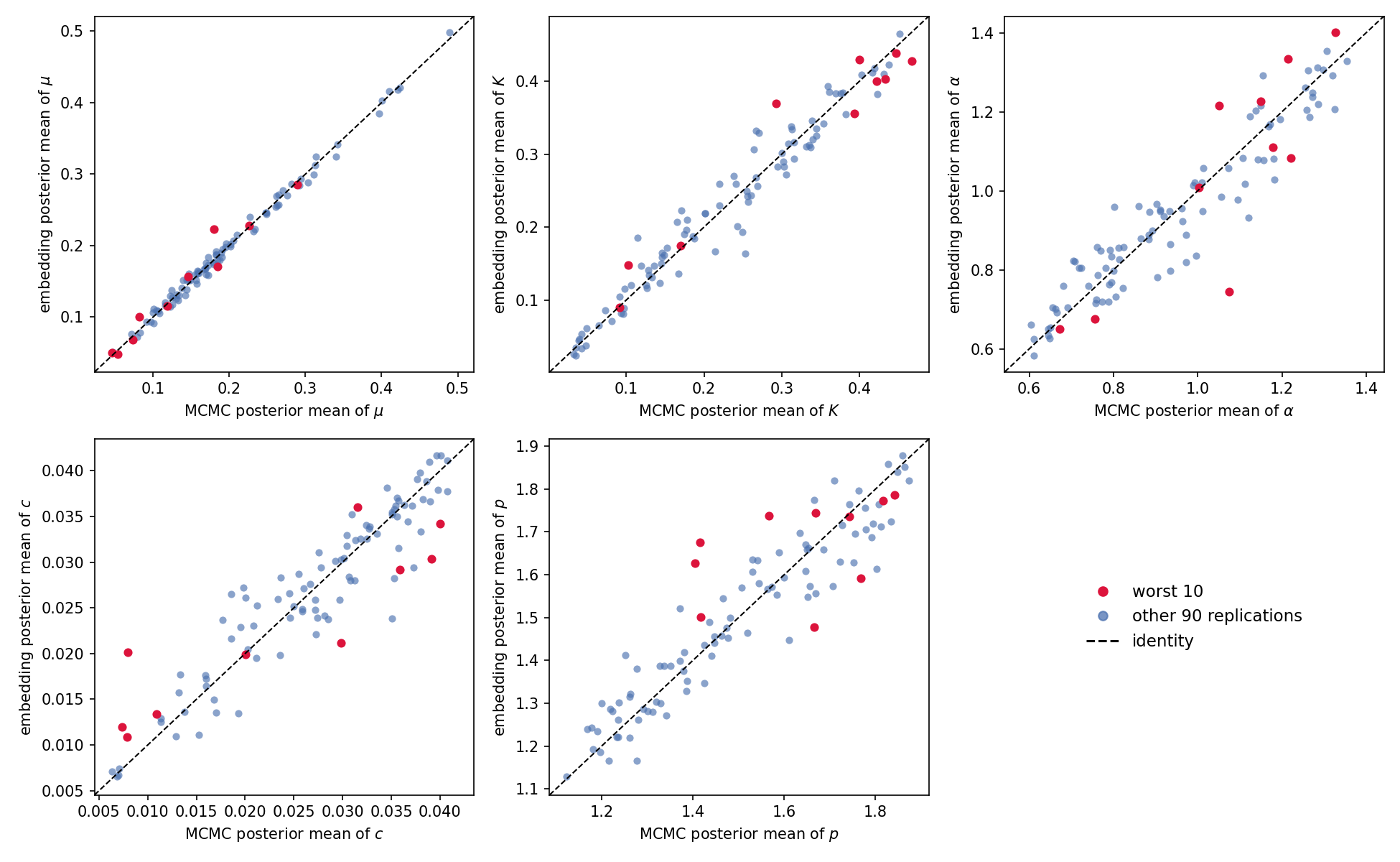}
\caption{Posterior means of the embedding pipeline against those of the MCMC
reference, one panel per parameter. Each dot is one of the \(100\)
replications, with the embedding pipeline's ten largest round-20 C2ST
accuracies in red; the dashed line is the identity.}
\label{fig:supp-mean-emb}
\end{figure}

%%%
%%% Figure S5
%%%

\begin{figure}[b!]
\centering
\includegraphics[width=0.85\linewidth]{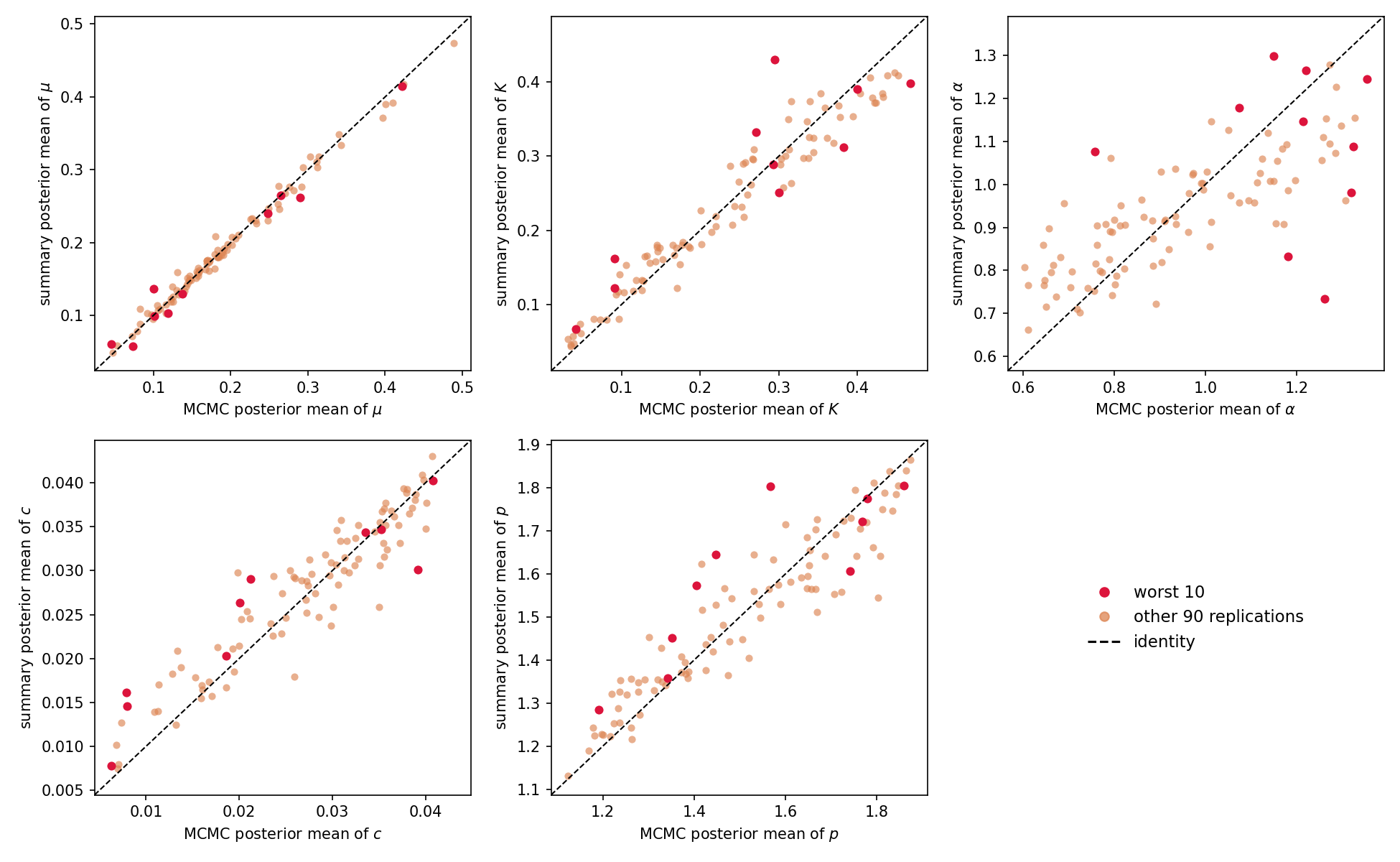}
\caption{As Figure~\ref{fig:supp-mean-emb}, for the summary pipeline.}
\label{fig:supp-mean-sum}
\end{figure}

%%%
%%% Figure S6
%%%

\begin{figure}[!t]
\centering
\includegraphics[width=0.85\linewidth]{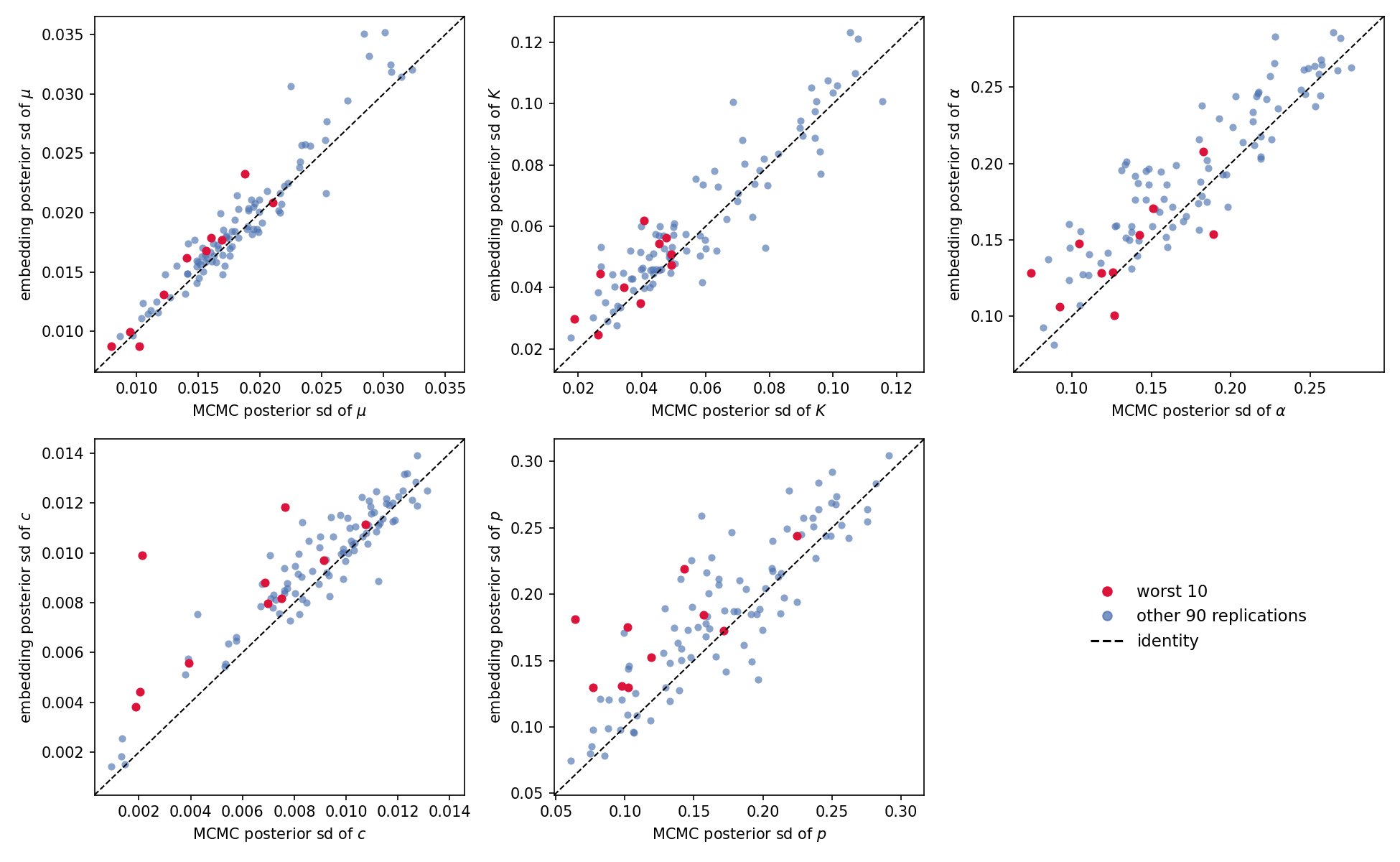}
\caption{Posterior standard deviations of the embedding pipeline against those
of the MCMC reference, one panel per parameter. Each dot is one of the \(100\)
replications, with the embedding pipeline's ten largest round-20 C2ST
accuracies in red; the dashed line is the identity, so points above it are
wider than the reference.}
\label{fig:supp-sd-emb}
\end{figure}

%%%
%%% Figure S7
%%%

\begin{figure}[!b]
\centering
\includegraphics[width=0.85\linewidth]{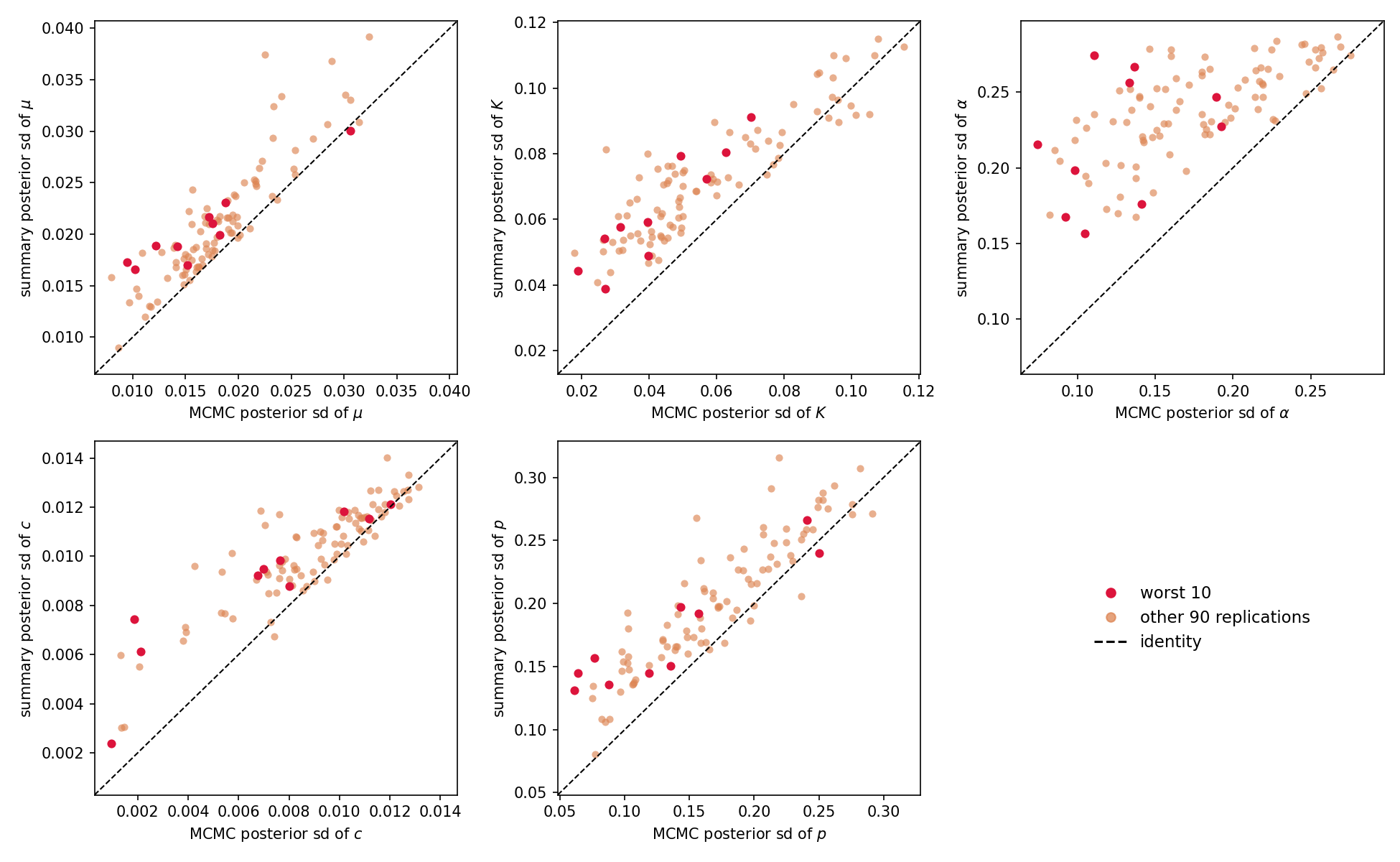}
\caption{As Figure~\ref{fig:supp-sd-emb}, for the summary pipeline.}
\label{fig:supp-sd-sum}
\end{figure}

%%%
%%% Figure S8
%%%

\begin{figure}[!t]
\centering
\includegraphics[width=0.9\linewidth]{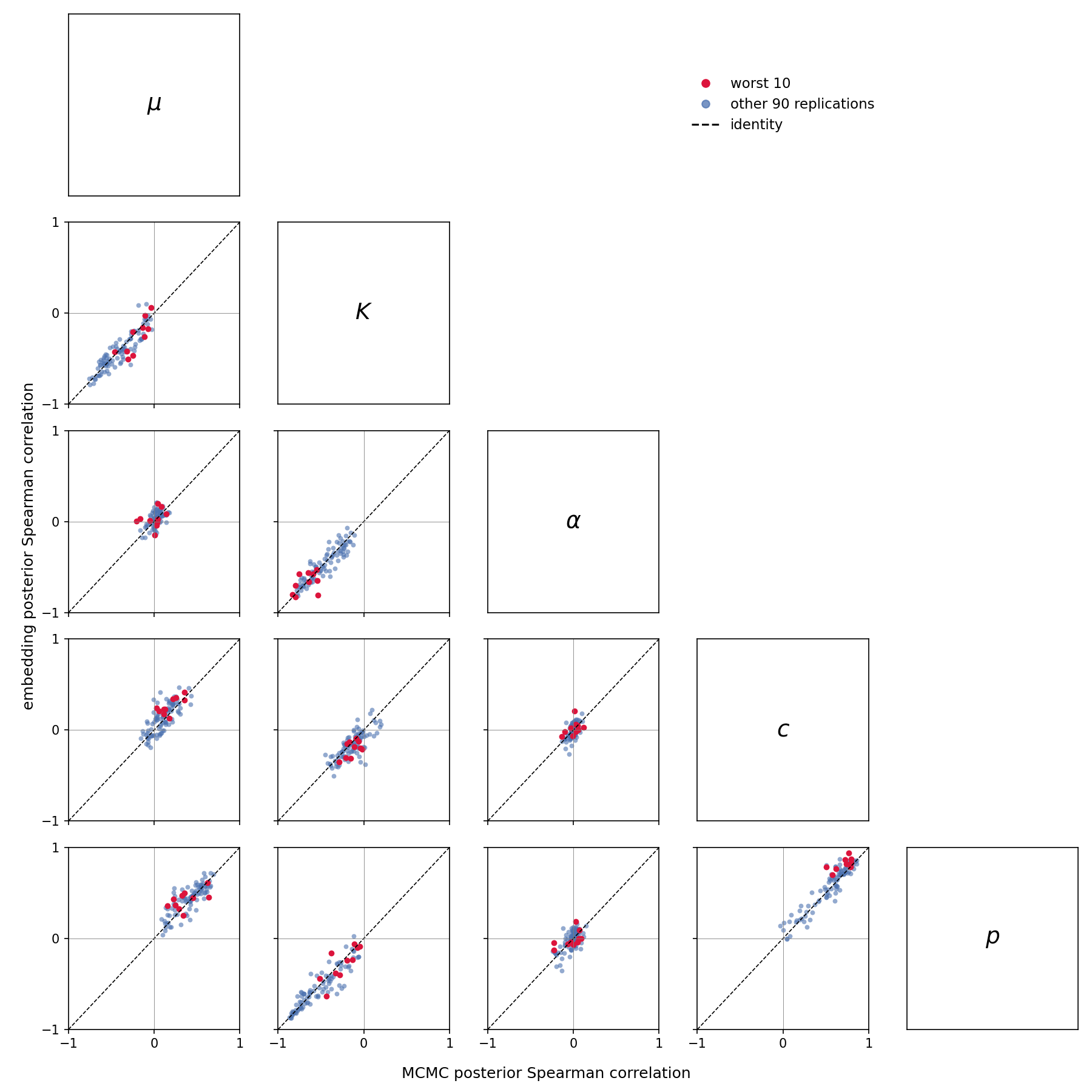}
\caption{Pairwise posterior correlations of the embedding pipeline against those
of the MCMC reference. Each panel of the lower triangle corresponds to one pair
of parameters, named on the diagonal; each dot is one of the \(100\)
replications, with the embedding pipeline's ten largest round-20 C2ST
accuracies in red, and the dashed line is the identity.}
\label{fig:supp-corr-emb}
\end{figure}

%%%
%%% Figure S9
%%%

\begin{figure}[!t]
\centering
\includegraphics[width=0.9\linewidth]{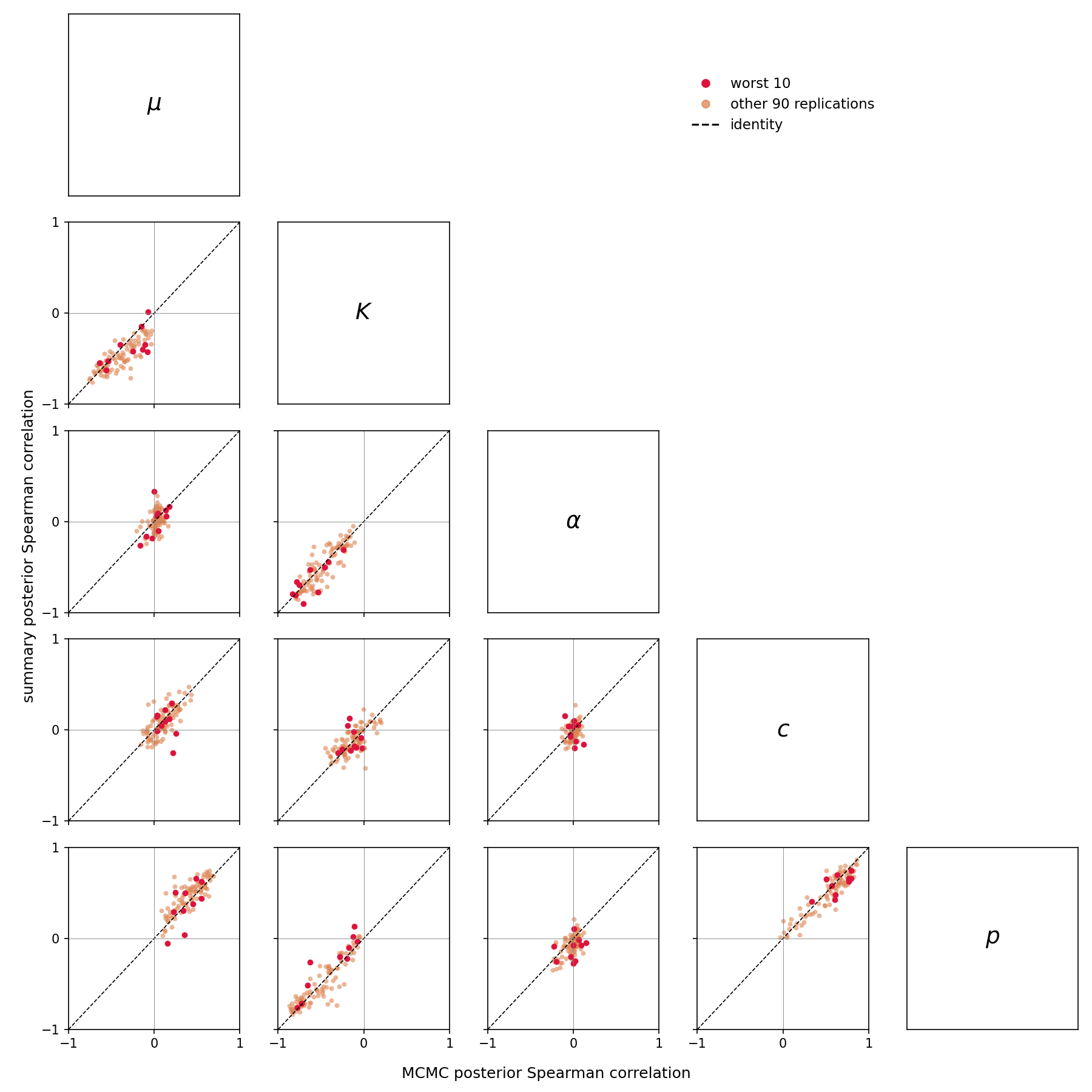}
\caption{As Figure~\ref{fig:supp-corr-emb}, for the summary pipeline.}
\label{fig:supp-corr-sum}
\end{figure}

\clearpage

\subsection{Comparison of marginal posteriors}
\label{subsec:supp-marginals}

\begin{figure}[H]
\centering
\includegraphics[width=\linewidth]{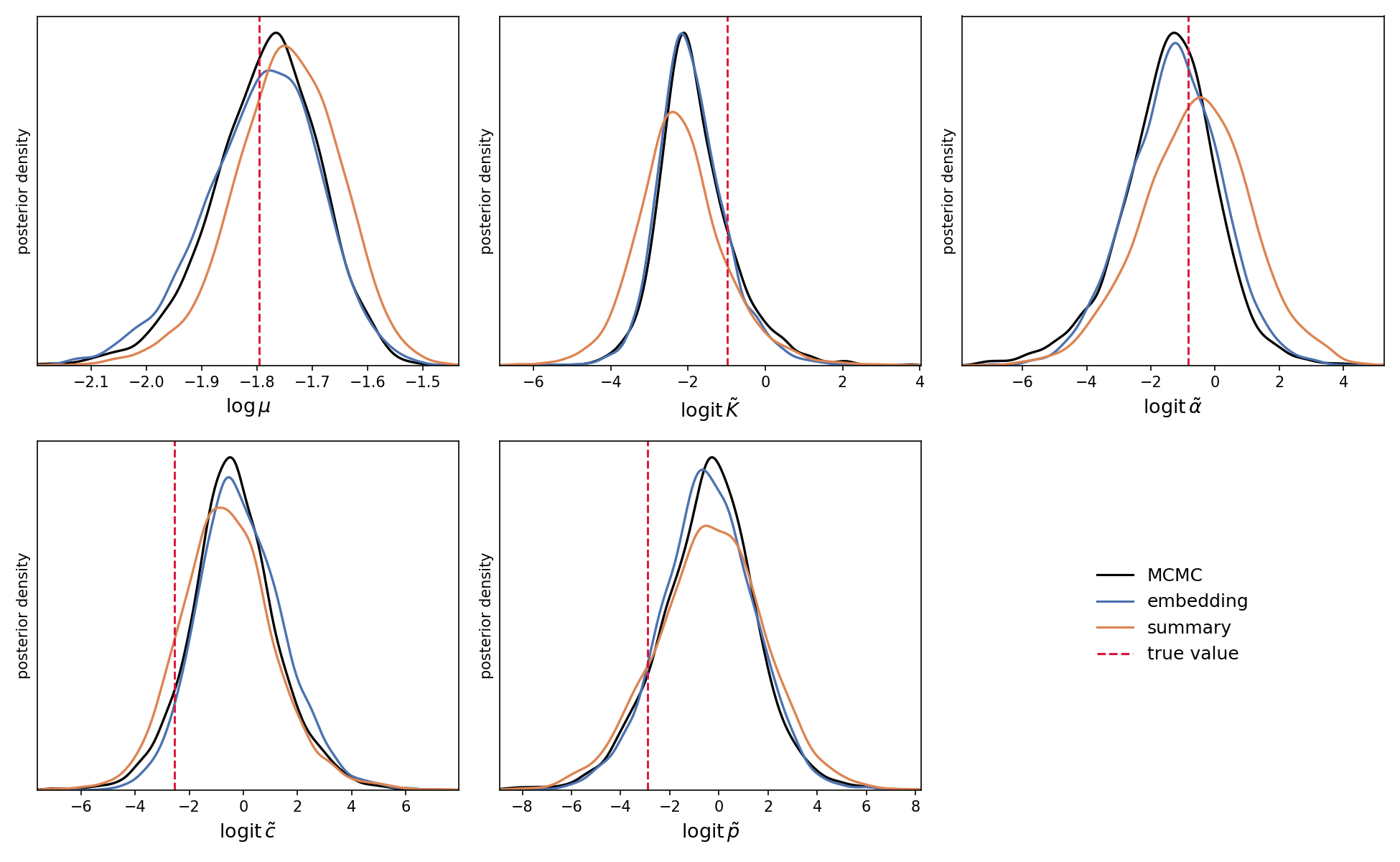}
\caption{Marginal posterior densities for seed 20 (\(n^\ast=0.22\), \(n=190\)) at the final (20th) training round. The C2ST accuracies are \(0.579\) for the embedding pipeline and \(0.655\) for the summary pipeline. Densities are shown on the unconstrained parameter scale used to train the NPE models: \(\log\mu\) and \(\operatorname{logit}(\tilde\theta)\), where \(\tilde\theta=(\theta-a)/(b-a)\) for parameters with a uniform prior
\(\mathcal{U}(a,b)\). MCMC draws are transformed to the same scale, and dashed lines indicate the true parameter values.}
\label{fig:supp-marginals}
\end{figure}

Figure~\ref{fig:supp-marginals} provides a representative comparison of the
marginal posterior densities from the two NPE pipelines with the MCMC reference
across all five parameters.

\end{document}